\documentclass[aps,prd,twocolumn,superscriptaddress]{revtex4-1}

\usepackage{longtable}
\usepackage{grffile} 
\usepackage{graphicx} 
\usepackage{amsmath}    
\usepackage{hyperref}   
\usepackage{subfigure}  

\newcommand{\pt}{\mbox{$p_T$ }}
\newcommand{\kt}{\mbox{$k_T$ }}

\newcommand{\ep}{\mbox{$e+p$}}
\newcommand{\eA}{\mbox{$e+$A}}
\newcommand{\eAu}{\mbox{$e+$Au}}
\newcommand{\dA}{\mbox{$d+$Au}}
\newcommand{\pA}{\mbox{$p+$A}}
\newcommand{ \pion }{ $\pi^{0}$ }

\newcommand{ \gammaA }{ \gamma^{*}+A\rightarrow h_{1}+h_{2}+X }
\newcommand{ \phase }{ dz_{h1}dz_{h2}d^{2}p_{h1\perp}d^{2}p_{h2\perp}  }

\begin{document}


\title{Probing Gluon Saturation through Dihadron Correlations \\ at an Electron-Ion Collider}

\author{L. Zheng}\email{lzheng@bnl.gov}
\affiliation{Key Laboratory of Quark and Lepton Physics (MOE) and \break Institute
of Particle Physics, Central China Normal University,\break Wuhan 430079, China}
\affiliation{Physics Department, Brookhaven National Laboratory, \break Upton, NY 11973, U.S.A.}
\author{E.C. Aschenauer}\email{elke@bnl.gov}
\affiliation{Physics Department, Brookhaven National Laboratory, \break Upton, NY 11973, U.S.A.}
\author{J.H. Lee}\email{jhlee@bnl.gov}
\affiliation{Physics Department, Brookhaven National Laboratory, \break Upton, NY 11973, U.S.A.}
\author{Bo-Wen Xiao}\email{xiaobowen@phy.ccnu.edu.cn}
\affiliation{Key Laboratory of Quark and Lepton Physics (MOE) and \break Institute
of Particle Physics, Central China Normal University,\break Wuhan 430079, China}

\date{\today}

\begin{abstract}
Two-particle azimuthal angle correlations have been proposed to be one of the
most direct and sensitive probes to access the underlying gluon dynamics
involved in hard scatterings. In anticipation of an Electron-Ion Collider (EIC),
detailed studies of dihadron correlation measurements in electron-proton and
electron-ion collisions at an EIC have been performed. The impact of such
measurements on the understanding of the different gluon distribution functions,
as a clean signature for gluon saturation and to constrain saturation models further,
has been explored. It is shown that dihadron correlation measurements will
be one of the key methods to probe gluon saturation phenomena at a future EIC.
\end{abstract}


\maketitle


\section{Introduction}

It is well-known that the concept of parton distribution functions (PDFs) and
their linear evolution through the Dokshitzer-Gribov-Lipatov-Altarelli-Parisi (DGLAP)
formalism~\cite{Dokshitzer:1977sg,Gribov:1972ri,Altarelli:1977zs} is a key ingredient for
perturbative quantum chromodynamics (pQCD) and has been successfully applied to
understand many hard processes in high energy hadronic collisions. Using
extractions of PDFs from pQCD fits to cross section data~\cite{Aaron:2009aa}, a large
number of high-energy physics experiments have established that
parton distributions, especially the gluon distribution, grow rapidly as the partonic
longitudinal momentum fraction, $x$, gets smaller. The high-energy
evolution of gluon density at fixed momentum transfer, $Q^2$, is described by the
so-called Balitsky-Fadin-Kuraev-Lipatov (BFKL) evolution
equation~\cite{Balitsky:1978ic}. The BFKL equation is a linear evolution equation
and it governs the evolution of the gluon distribution with respect to $x$. Its
solution exhibits a rapid increase as $x$ gets smaller. However, the gluon
density cannot grow arbitrarily large, since this would violate the unitarity
limit for forward scattering amplitudes or the Froissart bound for total cross
sections at very high energies. Recent experimental data at rather small $x$
have provided us some intriguing evidence for the existence of a novel QCD
regime, namely the saturation regime, which cannot be fully described by
linear QCD evolution approaches~\cite{Stasto:2000er,Armesto:2004ud,Gelis:2010nm}.

The idea of saturation physics can be briefly described as follows: when the
gluon density at low $x$ becomes so large that different gluon clouds with fixed
transverse size $\sim 1/Q^{2}$ start to overlap with each other, the QCD
evolution dynamics essentially becomes nonlinear~\cite{Gribov:1984tu,Mueller:1985wy}. 
It is conceivable that gluons can recombine in a dense medium, thereby taming further
rapid growth of the gluon density. The non-linear extension of the BFKL
equation is given by the Balitsky-Kovchegov equation~\cite{Balitsky:1995ub}.
This non-linear dynamical effect can be enhanced with a nuclear target, where the
interaction develops over a longitudinal distance of the order of the nuclear size
or larger. In this case the nucleons located at the same impact factor
cannot be distinguished from each other. Gluons from different nucleons can
amplify the total transverse gluon density by a factor of $A^{1/3}$ for a
nucleus with mass number $A$. Typically, a characteristic scale $Q_{s}(x,A)$ can
be introduced to describe the transition to the saturation region. For
$Q^{2}>Q_{s}^{2}$, the target hadron is usually treated as a dilute system,
whereas $Q^{2}<Q_{s}^{2}$ corresponds to the case with a highly dense saturated
hadron with a large parton density. Therefore, one can define a boundary with
$Q^2=Q_s^2(x)$ in the $x\textrm{-}Q^2$ plane to describe the transition from the
non-linear saturation regime to the linear dilute regime. The main physical
ingredient of the saturation formalism is to incorporate the unitarity
constraint for high-energy scattering amplitudes through the inclusion of
non-linear recombination in the quantum evolution of hadronic wave functions.

Although it is not conclusive that such a saturated regime has been discovered
at presently running high-energy experimental facilities, it has certainly
received a great amount of theoretical support with the development of the Color
Glass Condensate (CGC) effective field
theory~\cite{McLerran:1993ni,Iancu:2002xk}. To achieve an accurate description
of experimental data it is essential to have theoretical calculations beyond
leading order, especially as the higher-order corrections are known to be
sizeable. This analysis accounts for gluon radiation in the calculation of the
dihadron cross section through the inclusion of Sudakov factors. It is one of
the most important corrections computed on one-loop level. This is vital for the
comparison between the theory calculation and the future Electron-Ion Collider
(EIC)~\cite{Boer:2011fh,Accardi:2012qut} data on dihadron correlations.

It has been a challenging task to constrain to what extent saturation is present
in available experimental data~\cite{Albacete:2010rh}. Azimuthal dihadron
correlations are considered to be a very compelling measurement to tell whether
the partonic system under study has reached the saturation regime or
not~\cite{Kharzeev:2004bw}. The azimuthal angle $(\Delta\phi)$ distribution of
correlated high-\pt hadron pairs uncovers the underlying jet properties on a
statistical basis. The near-side peak ($\Delta\phi=0$) of this $\Delta\phi$
distribution is dominated by the fragmentation from the leading jet, while the
away-side peak ($\Delta\phi=\pi$) is expected to be dominated by back-to-back
jets produced in the hard $2\rightarrow2$ scattering. At sufficiently high
parton densities, when saturation effects dominate, incoming gluons normally
carry a typical transverse momentum at a scale of $Q_{s}>Q$, which significantly
increases the transverse momentum imbalance of the back-to-back jets. As a
result, saturated gluons from the target tend to smear the back-to-back picture
and suppress the away-side peak in the $\Delta\phi$ distribution.

The observed suppressions in dihadron correlation measurements at forward
rapidities performed in \dA\ $\sqrt{s}=200$ GeV collisions at
RHIC~\cite{Arsene:2004ux,Adams:2006uz,Adare:2011sc,Braidot:2010ig,Li:2012bn} are
perhaps the most suggestive evidence of the onset of the
saturation regime in present data. Rather significant suppression of the
away-side correlation is observed, when one compares the data for central \dA\
collisions to peripheral \dA\ collisions at forward rapidities. The
qualitative feature of this suppression was first predicted by
Marquet~\cite{Marquet:2007vb} based on saturation physics/CGC calculations. The
strength of back-to-back correlations and the depletion of the away-side peak
measured in these experiments can be quantitatively described in the saturation
formalism~\cite{Albacete:2010pg, Stasto:2011ru, Lappi:2012nh}.

At an EIC, we can access dihadron correlations in deep inelastic scattering
(DIS) data from \eA\ and \ep\ collisions, which can provide a clean and
well-controlled signature of saturation physics complementary to the current \dA\
or \pA\ measurements. They also provide the opportunity to study a fundamental
gluon distribution that cannot be accessed today. It has been shown in the
recent theoretical development of small $x$ physics that there are two different
unintegrated gluon distributions (UGDs); namely the Weizs\"{a}cker-Williams (WW)
gluon distribution and the dipole gluon distribution, which are involved in the
calculation of various observables~\cite{Dominguez:2010xd}. Since all other
gluon distributions appearing in various processes can be constructed from these
two UGDs in the large $N_c$ limit of QCD, they can be considered as the
universal and fundamental building blocks for all UGDs. Furthermore, the WW
gluon distribution can be interpreted as the gluon density in the light cone
gauge, while the dipole gluon distribution has no such probabilistic
interpretation. In addition, we want to emphasize that the WW gluon distribution
only appears in few physical processes exclusively, and currently there is very
little knowledge about its behavior. Fortunately, the WW gluon distribution is
the only UGD involved in the DIS dijet process~\cite{Dominguez:2011wm}, which
provides us a unique and clean means to measure the WW gluon distribution.

In this paper, based on the most recent theoretical developments in saturation
physics, we perform a detailed study of the feasibility, expected precision, and
physics impact of dihadron correlation measurements on gluon dynamics in the
small $x$ region at a future high-luminosity, high-energy EIC. We will
demonstrate that, at a future EIC such as eRHIC at BNL or MEIC at
JLab (see Sec.5 of Ref.~\cite{Accardi:2012qut} and references therein for more
details), it is feasible to perform the discussed dihadron correlation measurement even
with a moderate integrated luminosity of $\mathcal{L}=1 \, fb^{-1}$. We present results
for two lepton-nucleus beam energy configurations, 10 GeV $\times$ 100 GeV and
20 GeV $\times$ 100 GeV, and compare the results for proton and gold beams. We use
pseudo-data generated by the Monte Carlo generator
PYTHIA~\cite{Sjostrand:2006za} integrated with nuclear PDFs, geometry and energy
loss to obtain a non-saturation baseline. The framework of
Ref.~\cite{Dominguez:2011wm} is used to obtain numerical predictions and to
study the size of the suppression of dihadron correlations in a saturation
formalism. Table~\ref{tab:varDef} shows the definitions of the kinematic
variables used in this study.

\begin{longtable*}[H]{ll}
\caption{ Kinematic variables \label{tab:varDef} } \\ \hline \hline
$q=(E_{e}-E_{e}^{'},\vec{l}-\vec{l^{'}})$ & 4-momentum of virtual photon \\
$Q^{2}=-q^{2}$	& Virtuality of exchanged photon  \\
$x_{Bj}=\frac{Q^{2}}{2P\cdot q}$ & Bjorken $x$, momentum fraction of the incoming nucleon taken by the struck quark in the electron rest frame\\
$y=\frac{P\cdot q}{P\cdot l}$	& Energy fraction of virtual photon with respect to incoming electron \\
$\sqrt{s}$	& Center of mass energy \\
$x_{g}$	& Longitudinal momentum fraction of gluon involved in hard interactions \\
$z_{h,q}=\frac{P\cdot P_{h,q}}{P\cdot q}$	& Energy fraction of of a hadron or quark with respect to virtual photon in target rest frame \\
$p_{T}$ & Transverse momentum of final state hadron with respect to virtual photon \\
$\Delta\phi$ & Azimuthal angle difference of the trigger $\vec{p_{T}}^{trig}$ and associate $\vec{p_{T}}^{assoc}$ \\
$\eta =-\ln\tan(\theta/2)$ & Pseudorapidity of the particles in lab frame \\
$q_\perp$ & Initial transverse momentum of gluons in $\gamma^{*}p$ center of mass frame \\
$k_{1\perp}, k_{2\perp}$ &	Transverse momentum of the dijets in $\gamma^{*}p$ center of mass frame \\
$p_{h1\perp}, p_{h2\perp}$ &	Transverse momentum of the trigger/associate particle in $\gamma^{*}p$ center of mass frame \\
$\hat{p_{T}}$ & Transverse momentum of final state partons in the center of mass frame of hard interaction \\
$k_{T}$	& Intrinsic transverse momentum of partons in the nucleon \\
$p_{T}^{\textrm{frag}}$ & Transverse momentum with respect to jet direction from hadronization \\
$Q_{s}$	&  Saturation scale \\ \hline \hline
\end{longtable*}

The rest of this article is organized as follows: in Sec.~\ref{sec:theory}, we
discuss the theoretical framework used for the prediction of saturation effects
in the dihadron correlation measurement. A brief comparison of dihadron
correlations in \eA\ versus \pA\ is provided in Sec.~\ref{sec:eAvspA}. In
Sec.~\ref{sec:MC}, we give an overview of the planned EIC project and present
simulation results for dihadron correlations at an EIC. Finally, we summarize
and conclude in Sec.~\ref{sec:summary}.

\section{Dihadron Correlations in the Saturation Formalism}\label{sec:theory}

According to the effective small-$x$ $k_t$ factorization established in
Ref.~\cite{Dominguez:2011wm}, which is briefly summarized above, the
back-to-back correlation limit of the dihadron production cross section can be used
to directly probe the WW gluon distribution $xG^{(1)}(x,q_{\perp})$. As a
comparison, the hadron production in semi-inclusive deep inelastic scattering
(SIDIS), as shown in Ref.~\cite{Marquet:2009ca}, is related to the so-called
dipole gluon distributions $xG^{(2)}(x,q_{\perp})$.

The coincidence probability $C(\Delta\phi)=\frac{N_{pair}(\Delta\phi)}{N_{trig}}$ is a
commonly exploited observable in dihadron correlation studies, in which
$N_{pair}(\Delta\phi)$ is the yield of the correlated trigger and associate
particle pairs, while $N_{trig}$ is the trigger particle yield. This
correlation function $C(\Delta\phi)$ depends on the azimuthal angle difference $\Delta\phi$
between the trigger and associate particles. In terms of theoretical
calculation, the correlation function is defined as
\begin{eqnarray} 
C(\Delta\phi) 
= & \frac{1}{\frac{d\sigma^{\gamma^{*}+A\rightarrow h_{1}+X}_{\textrm{SIDIS}}}{dz_{h1}}}
\frac{d\sigma^{\gammaA}_{\textrm{tot}}}{dz_{h1}dz_{h2}d\Delta\phi}.
\label{eqn:cdphiCGC} 
\end{eqnarray}

Let us consider a process of a virtual photon scattering on a dense nuclear target
producing two final state back-to-back $q\bar q$ jets:
$\gamma^{*}+A\rightarrow q(k_{1})+\bar{q}(k_{2})+X$, in which $k_1$ and $k_2$ are
the four momenta of the two outgoing quarks. This process is the
dominant one in the low-$x$ region, since the gluon distribution is much larger than
the quark distributions inside a hadron at high energy. The back-to-back correlation
limit indicates that the transverse momentum imbalance is much smaller than each
individual momentum: $q_{\perp}=|k_{1\perp}+k_{2\perp}|\ll P_{\perp}$, with
$P_{\perp}$ defined as $(k_{1\perp}-k_{2\perp})/2$. At leading order (LO), the
dihadron total cross section, which includes both the longitudinal and
transverse contributions, can be written as follows~\cite{Dominguez:2011wm}:

\begin{widetext}
\begin{eqnarray}\label{eqn:xspairCGC}
\frac{d\sigma^{\gammaA}_{\textrm{tot}}}{\phase} = & C\int^{1-z_{h2}}_{z_{h1}} dz_{q}
\frac{z_{q}(1-z_{q})}{z^{2}_{h2}z^{2}_{h1}}
d^{2}p_{1\perp}d^{2}p_{2\perp}\mathcal{F}(x_{g},q_{\perp})
\mathcal{H}_{\textrm{tot}}(z_{q},k_{1\perp},k_{2\perp}) \\ \nonumber
& \times \sum_{q}e^{2}_{q}D_{q}(\frac{z_{h1}}{z_{q}},p_{1\perp})
D_{\bar{q}}(\frac{z_{h2}}{1-z_{q}},p_{2\perp}),
\end{eqnarray}  
\end{widetext}
where $C=\frac{S_{\perp}N_{c}\alpha_{em}}{2\pi^{2}}$ gives the normalization
factor, with $S_{\perp}$ being the transverse area of the target, $z_{q}$ is the
longitudinal momentum fraction of the produced quark with respect to the incoming
virtual photon, $\mathcal{H}_\textrm{tot}$ is the combined hard factor,
$k_{1\perp}$ and $k_{2\perp}$ are the transverse momenta of the two quarks, while
$p_{h1\perp}$ and $p_{h2\perp}$ are the transverse momenta of the two corresponding
produced hadrons respectively. $\mathcal{F}(x_{g},q_{\perp})$ comes from the
relevant WW gluon distribution $xG^{(1)}(x_g,q_\perp)$ evaluated with the gauge
links for a large nucleus at small $x$ by using the McLerran-Venugopalan
model~\cite{McLerran:1993ni},
\begin{equation} 
\mathcal{F}(x_{g}, q_{\perp}) =
\frac{1}{2\pi^{2}} \int d^{2}r_{\perp} e^{-iq_{\perp}r_{\perp}}
\frac{1}{r^{2}_{\perp}}[1-\exp(-\frac{1}{4}r^{2}_{\perp}Q^{2}_{s})],
\end{equation}
in which
$x_{g}=\frac{z_{q}p_{h1\perp}^2}{z_{h1}^2s}+\frac{(1-z_{q})p_{h2\perp}^2}{z_{h2}^2s}+\frac{Q^2}{s}$ is the longitudinal momentum fraction of the small-$x$ gluon with respect to the target hadron and $Q_{s}$ is the gluon saturation scale. $D_{q}(\frac{z_{h}}{z_{q}},p_{\perp})$ represents the transverse momentum dependent fragmentation functions, where $p_{\perp}$ shows the additional transverse momentum introduced by fragmentation processes. There can be more sophisticated model description of the WW gluon distribution, which involves a numerical solution to the BK type evolution for the WW gluon distribution~\cite{Dominguez:2011gc,Dumitru:2011vk}. But studying the impact of these PDFs is beyond the scope of this work presented here.

In principle, the so-called linearly polarised gluon
distribution~\cite{Metz:2011wb,Dominguez:2011br} also contributes to the dihadron
correlation and can be systematically taken into account. This part of the
contribution comes from an averaged quantum interference between a scattering
amplitude and a complex conjugate amplitude with active gluons linearly
polarized in two orthogonal directions in the azimuthal plane. Numerical
calculation shows that this contribution is negligible for dihadron back-to-back
correlations. Also, this type of contribution vanishes when the dihadron correlation function 
is averaged over the azimuthal angle of the trigger particle.

As to the single-inclusive-hadron production cross section, which enters the
denominator of the definition of the correlation function $C(\Delta\phi)$, it
can be calculated from the saturation physics/CGC formalism~\cite{Marquet:2009ca} as
follows:
\begin{widetext}
\begin{eqnarray}\label{eqn:xstrigCGC}
\frac{d\sigma^{\gamma^{*}+A\rightarrow h_{1}+X}_{\textrm{SIDIS}}}{dz_{h1}d^{2}p_{h1\perp}} 
= & C \int^{1}_{z_{h1}}dz_{q} \int
d^{2}q_{\perp}F_{x_{g}}(q_{\perp})H_{\textrm{SIDIS}}(k_{\perp},q_{\perp},Q) \\ \nonumber
  & \times \sum_{q} e^{2}_{q}\frac{z_{q}}{z^{2}_{h1}}D_{q}(\frac{z_{h1}}{z_{q}},p_{\perp}),
\end{eqnarray}
\end{widetext}
where $H_{\textrm{SIDIS}}$ is the $q_{\perp}$ dependent hard factor for SIDIS,
which includes both the longitudinal and transverse photon contribution. Here
$F_{x_{g}}(q_\perp)$, which is related to $xG^{(2)}(x_g,q_\perp)$, is the Fourier
transform of the dipole cross section:
\begin{eqnarray}
F_{x_{g}}(q_{\perp})& = &\int \frac{d^{2}r}{2\pi^{2}}e^{iq_{\perp}\cdot
r_{\perp}} \frac{1}{N_{c}} Tr\langle
U(r_{\perp})U^{\dag}(0)\rangle_{\rho} \\ \nonumber
& \simeq &\frac{1}{\pi
Q^{2}_{sA}}\exp[-\frac{q_{\perp}^{2}}{Q^{2}_{sA}}]. 
\end{eqnarray}
It has been suggested in Refs~\cite{Dominguez:2011gc, Dumitru:2011vk} that both
dipole and WW gluon distributions have similar geometric scaling behavior.
Therefore, one can parameterize these gluon distributions following the
Golec-Biernat W\"{u}sthoff (GBW)~\cite{GolecBiernat:1998js} model calculation, in
which $Q^{2}_{sA}(x)=c(b)A^{1/3}Q^{2}_{s0}(x/x_{0})^{-\lambda}$, with $Q_{s0}=1$
GeV, $x_{0}=3.04\times 10^{-4}$ and  $\lambda=0.288$. The gluon saturation
momentum is related to $Q^{2}_{sA}(x)$ by
$Q_s^2(x)=\frac{2N_c^2}{N_c^2-1}Q^{2}_{sA}(x)$. $c(b)=c(0)\sqrt{1-b^{2}/R^{2}}$
gives the nuclear profile dependence with a radius $R$, where $b$ is the impact
parameter. As it is not an easy task to determine the exact impact parameter in \eA\
collisions, a median number $c(b)=0.8$ is used for the estimation,
which is supposed to average the nucleus geometry effectively. The parametrized
DSS fragmentation function~\cite{deFlorian:2007aj},
$D(z,p_{\perp})=D(z)\frac{1}{\pi\langle
p^{2}_{\perp}\rangle}e^{\frac{-p_{\perp}^{2}}{\langle p^{2}_{\perp}\rangle}}$ with
$\langle p^{2}_{\perp}\rangle=0.2 \, \mathrm{GeV}^{2}$, is used to compute the hadron
production.

\begin{figure}
\includegraphics[width=0.5\textwidth]{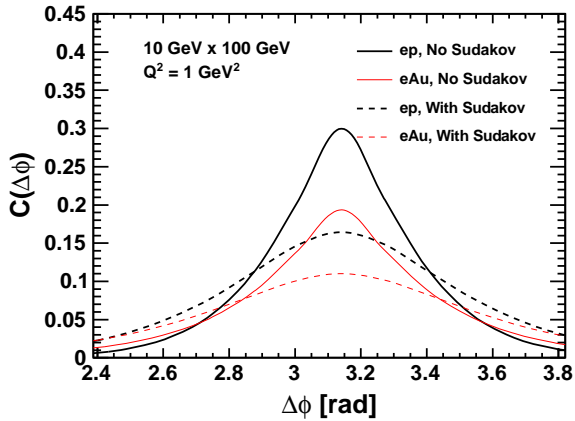} 
\caption[theory prediction of dihadron correlation with Sudakov] {[color online] 
$\pi^0$-correlation curves calculated in the saturation formalism at 10
GeV$\times $100 GeV for \ep\ (thick line) and \eAu\ (thin line) with (dashed
curve) and without (solid curve) the Sudakov factor. The kinematics chosen are
$y=0.7$, $Q^2=1 \, \textrm{GeV}^2$, $z_{h1}=z_{h2}=0.3$, $p_{h1\perp}>2 \,
\mathrm{GeV}/c,1 \, \mathrm{GeV}/c<p_{h2\perp}<p_{h1\perp}$.
\label{fig:dihadron_theory_sud}
}    
\end{figure}

By utilizing Eq.~(\ref{eqn:xspairCGC}) and Eq.~(\ref{eqn:xstrigCGC}), one can
straightforwardly calculate the coincidence probability. The theoretical
prediction at the Born level for the suppression of the away-side of the
dihadron correlation measurement is shown by the solid curves in
Fig.~\ref{fig:dihadron_theory_sud}.

All the above results are estimated based on the LO Born level
contribution. At the EIC energy scale the one-loop
contribution~\cite{Mueller:2012uf}, which is also known as the so-called Sudakov
factor, can be important as well. To include the Sudakov factor contribution at
leading double logarithm level, one can rewrite the relevant WW distribution as
follows~\cite{Mueller:2013wwa}:
\begin{widetext}
\begin{equation}
\mathcal{F}(x_{g}, q_{\perp}) =
\frac{1}{2\pi^{2}} \int d^{2}r_{\perp} e^{-iq_{\perp}r_{\perp}}
\frac{1}{r^{2}_{\perp}}[1-\exp(-\frac{1}{4}r^{2}_{\perp}Q^{2}_{s})]
\exp[-\frac{\alpha_sN_c}{4\pi}\ln^2\frac{K^2r_{\perp}^2}{c^2_0}],
\label{eqn:Sudakov}
\end{equation}
\end{widetext}
where $K^2$ represents the hard momentum scale in two-particle production
processes. It can be chosen as $K^2=P^2_{\perp}$ or $K^2=Q^2$, depending on which
one is larger, and $c_0=2e^{-\gamma_E}$ with the Euler constant $\gamma_E$. It
is known that the single logarithmic terms as well as the next-to-leading order
(NLO) contribution of the Sudakov factor also have sizeable contributions 
compared to the above leading double logarithmic contribution. Therefore, the
numerical value of $\alpha_s$ in the Sudakov factor used in this calculation may
be different from what one normally expects according to a QCD running coupling
constant calculation.

One needs to pay attention to the applicability of this calculation.
As the GBW model is not sufficient to describe the UGDs in the region where 
$q_{\perp}$ is much larger than $Q_s$, we should limit this calculation to the
saturation region ($x_g<0.01$) to ensure the GBW model
can be applied. Additionally, to ensure that the power corrections to the
two-particle production are negligible, one needs the magnitude of the jet
transverse momenta $P_{\perp}$ to be much larger than $Q_s$.

The current calculations are performed for $Q^2$ of the same order as
$P^2_{\perp}$. For pair production, the Sudakov factor is usually due to
a scale difference between $P_{\perp}$ and the dijet momentum imbalance $q_{\perp}$.
Because we have required that $P_{\perp}\gg q_{\perp}$ as discussed above, it is
necessary to include the Sudakov contribution.  As for the trigger hadron
inclusive cross section, the Sudakov factor is not important, since the trigger
hadron \pt is of the same order as $Q$ and $P_\perp$. An illustration of
this Sudakov effect with $\alpha_s=0.35$ can be found in
Fig.~\ref{fig:dihadron_theory_sud} labeled by the dashed lines.
It is worthwhile to point out that the Sudakov effect in a nuclear environment
is still not very well known. In the current small $x$ scenario as shown in
Eq.~(\ref{eqn:Sudakov}), it is convoluted with the gluon distribution function.
The theoretical calculation indicates that the Sudakov factor has no nuclear
$A$ dependence at LO. As shown in Fig.~\ref{fig:dihadron_theory_sud}, the away-side 
suppression of the dihadron correlation is due to the combination of the
Sudakov suppression and saturation effects. It is conceivable that the
suppression due to saturation effects shall become more and more dominant when
the ion beam species are changed from proton to gold, while the Sudakov effect
remains more or less the same. 


\section{Connections to \pA\ dihadron correlations}\label{sec:eAvspA}

Compared to existing \pA\ or \dA\ dihadron correlation data, there are
several advantages to measuring dihadron correlations in \eA\ collisions. 
One valuable feature is that one can make use of the scattered electron to
reconstruct kinematic information event by event. Measuring
the scattered electron allows us to model-independently determine the required
kinematic variables $x$ and $Q^{2}$, which is essential for probing the
underlying gluon dynamics precisely. Another advantage comes from the point-like
structure of electrons. Since electrons have no substructure and they couple to
virtual photons rather weakly, the probability to have multiple emission in \eA\
is very small compared to \pA. This kind of multiple emission or
interaction introduces a significant amount of uncorrelated two-particle production,
which is known as the ``pedestal'' effect in RHIC \dA\ collisions. To understand the
so-called pedestal effect, one needs to take into account the double parton
scattering, which includes two independent and uncorrelated hard scatterings. In
contrast, as explained above, the pedestal contribution should be negligible and
under control in the \eA\ dihadron correlation measurement.

\begin{figure}[hbt] 
\begin{center}
\includegraphics[width=0.5\textwidth]{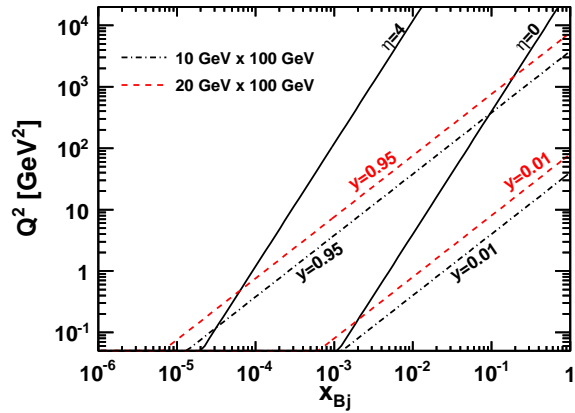} 
\end{center} 
\caption[kinematics coverage of eRHIC vs RHIC]
{[color online] The eRHIC kinematics coverage compared to \pA\ at RHIC. The dash-dotted and dashed
lines show the eRHIC kinematics for the beam energies of 10 GeV
$\times$100 GeV and 20 GeV $\times$100 GeV, respectively. The solid lines 
represent the RHIC coverage at $\sqrt{s}=200$ GeV for $\eta=0$ and $\eta=4$, where
$\eta=-\ln\tan(\theta/2)$ is the pseudorapidity of the particles.
\label{fig:eAvspA_kinematics}}
\end{figure}

In addition, as is shown in Fig~\ref{fig:eAvspA_kinematics}, the kinematic
coverage of the planned eRHIC realization of the EIC
is very similar to the measured RHIC \dA\ data and
extends to the desired small-$x$ region. Therefore, it will be a more precise and definite
measurement compared to what we already know from the RHIC \dA\ data.
Here, the RHIC kinematics lines are calculated with the assumption that the major fraction of 
the parton energy is mainly taken by the hadrons, which is not always true.

The measurement of dihadron correlations in \eA\ collisions is
interesting by itself. It provides us with a golden opportunity to directly
measure the saturated WW gluon distribution. Through detailed calculations,
Ref.~\cite{Dominguez:2011wm} summarizes the involvement of these two basic
gluon distributions in different observables. It is interesting
to note that the dipole gluon distribution function is involved in most
known processes, especially inclusive DIS measurements, which provides us with a
lot of the essential information of the dipole scattering amplitude. On the
other hand, the WW gluon distribution contributes to only a few of these
processes, thus very little knowledge about the WW distribution exists from
the current experimental data. In addition, unlike the dijet process in \pA,
which receives contribution from both the dipole and WW gluon distribution, the WW gluon
distribution is the only UGD that contributes to \eA\ Dijet production. Considering
that the WW gluon distribution can be physically interpreted as the number density
of gluons inside a nuclear wave function, while the dipole gluon distribution
does not have such interpretation, it is important and fundamental to
acquire direct information on the WW gluon distribution through dihadron
correlation measurements at an EIC.

\section{Electron-Ion Collider and Simulations}\label{sec:MC}

\subsection{The Electron-Ion Collider and its Detector}

Two independent designs for an EIC are being developed in the United States:
eRHIC, at Brookhaven National Laboratory (BNL); and MEIC/ELIC at Thomas
Jefferson National Laboratory (JLab).  The following studies will focus on the
eRHIC version of an EIC and the new model detector at eRHIC. The eRHIC design at
BNL reuses the available infrastructure and facilities of RHIC's high-energy
polarized proton/ion beam. A new electron beam, based on Energy Recovery LINAC
(ERL) technology, is to be built inside the current RHIC tunnel. At eRHIC, the
collision luminosity is expected to be in the order of $10^{33-34}
\textrm{cm}^{-2}\textrm{s}^{-1}$. The full range of proton/ion beam energies will be accessible
from the beginning of operations, while the electron beam energy will start with
$10-15$ GeV and later be increased to $20$ GeV.

The tracking system of the baseline eRHIC detector will consist of a TPC, GEM
and silicon detectors spanning a range of $-4<\eta<4$ in pseudorapidity. The
end-cap and barrel region on the detector will be equipped with electromagnetic
calorimeters covering $-4.5<\eta<4.5$. Hadronic calorimeter will be used mostly
for jet physics at full energy in the forward (hadron beam going direction) and
backward (electron beam going direction) rapidities spanning $2<|\eta|<4.5$.
Projected momentum and energy resolutions of these devices are better than a few
percent, which extends the capability of this detector to a large variety of
physics topics.

\begin{figure*} 
\begin{center}

\includegraphics[width=0.45\textwidth]
{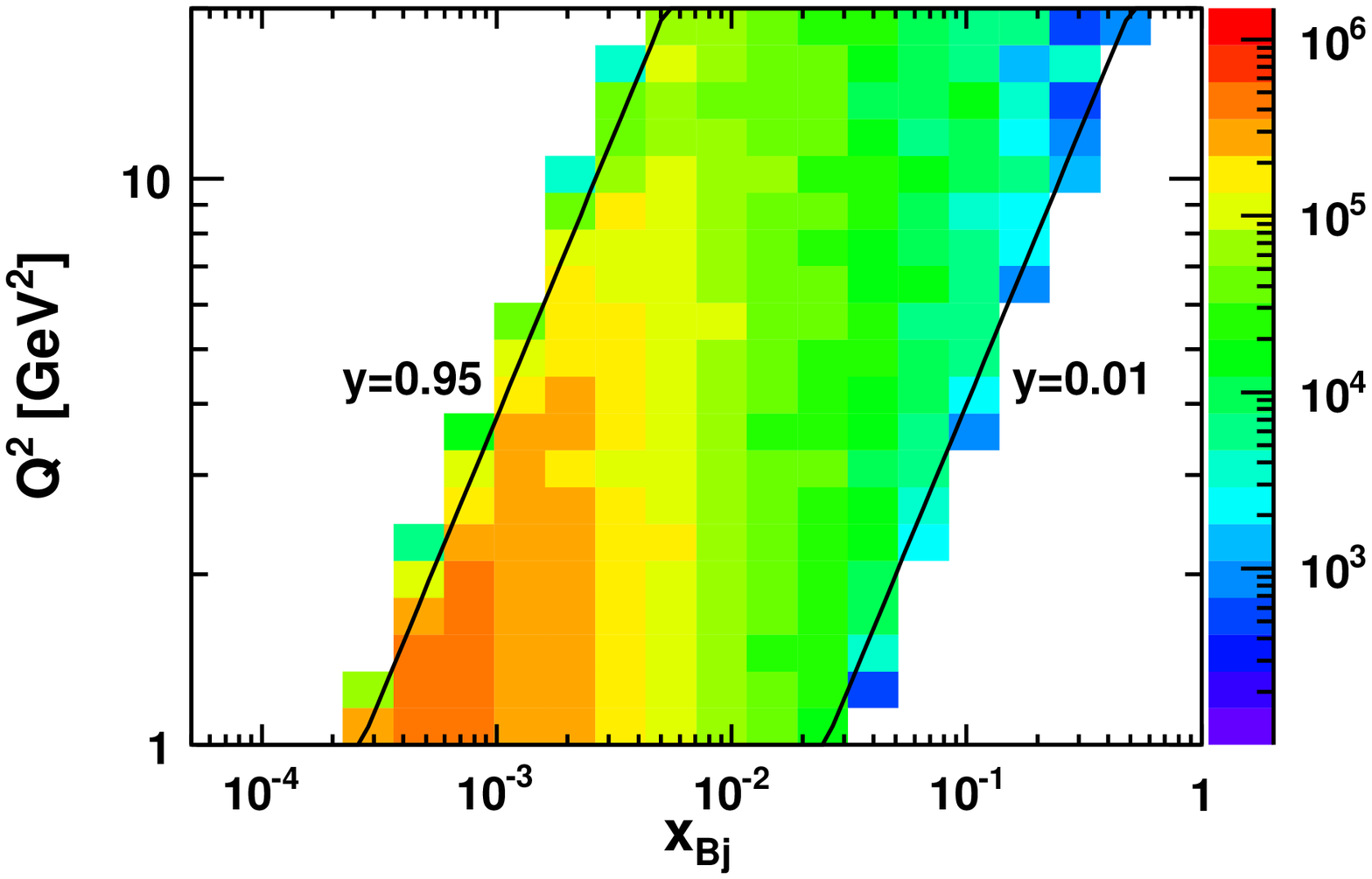}
\includegraphics[width=0.45\textwidth]
{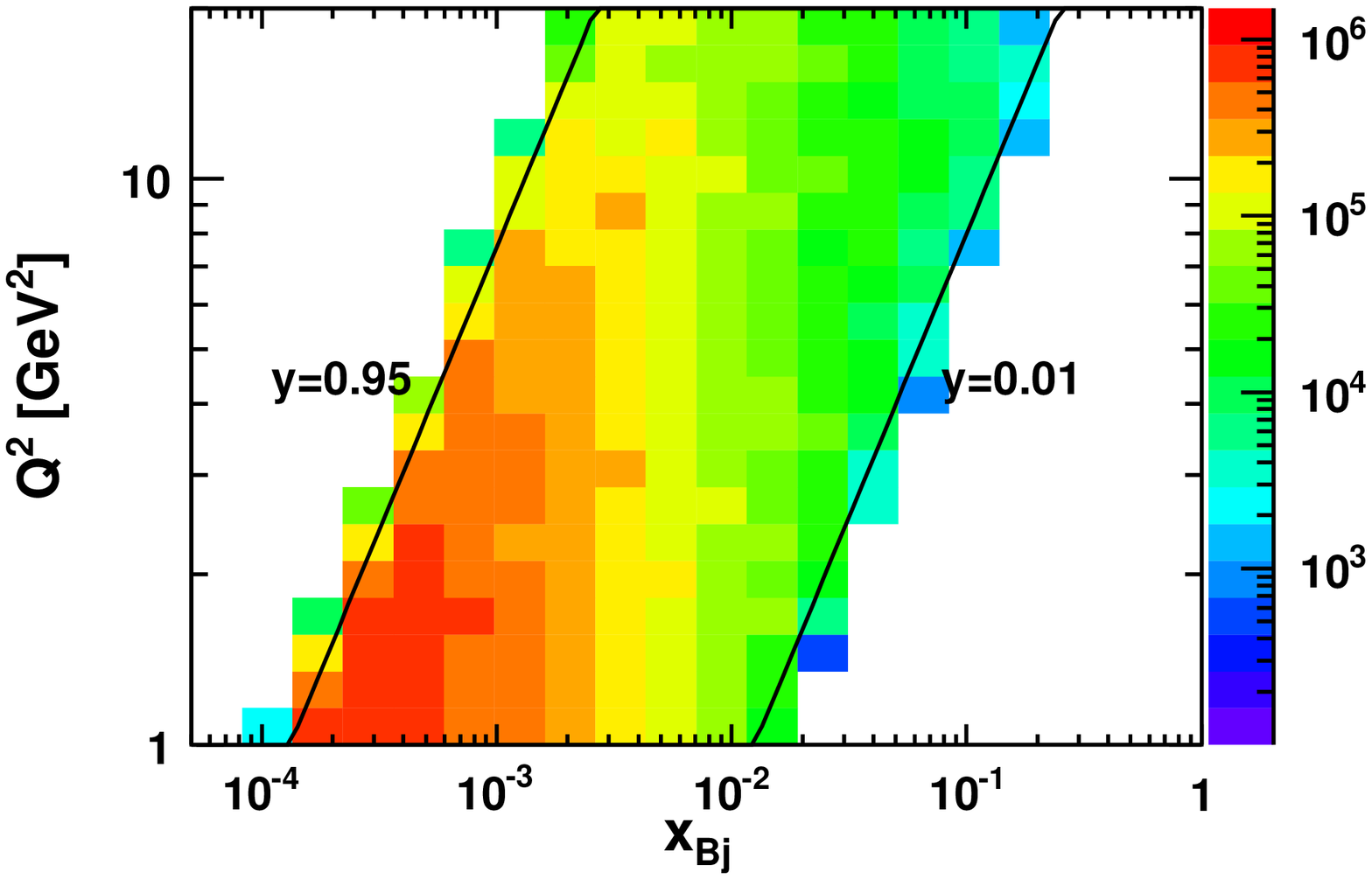}

\end{center} 
\caption[statistics in allowed kinematics region]{[color online] Expected yields of charged
particle pairs at transverse momentum $p_{T}>1$ GeV/$c$ in bins of 
($Q^{2}$, $x_{Bj}$) for an integrated luminosity of $1 \, fb^{-1}$ for 
\ep\ 10 GeV $\times$100 GeV (Left) and 20 GeV $\times$100 GeV (Right) in the kinematic
range of $1\, \textrm{GeV}^{2}<Q^{2}<20 \, \textrm{GeV}^{2}$, and $0.01<y<0.95$.}
\label{fig:Q2VsxBj}
\end{figure*}

\begin{figure*} 
\begin{center} 

\includegraphics[width=0.45\textwidth]
{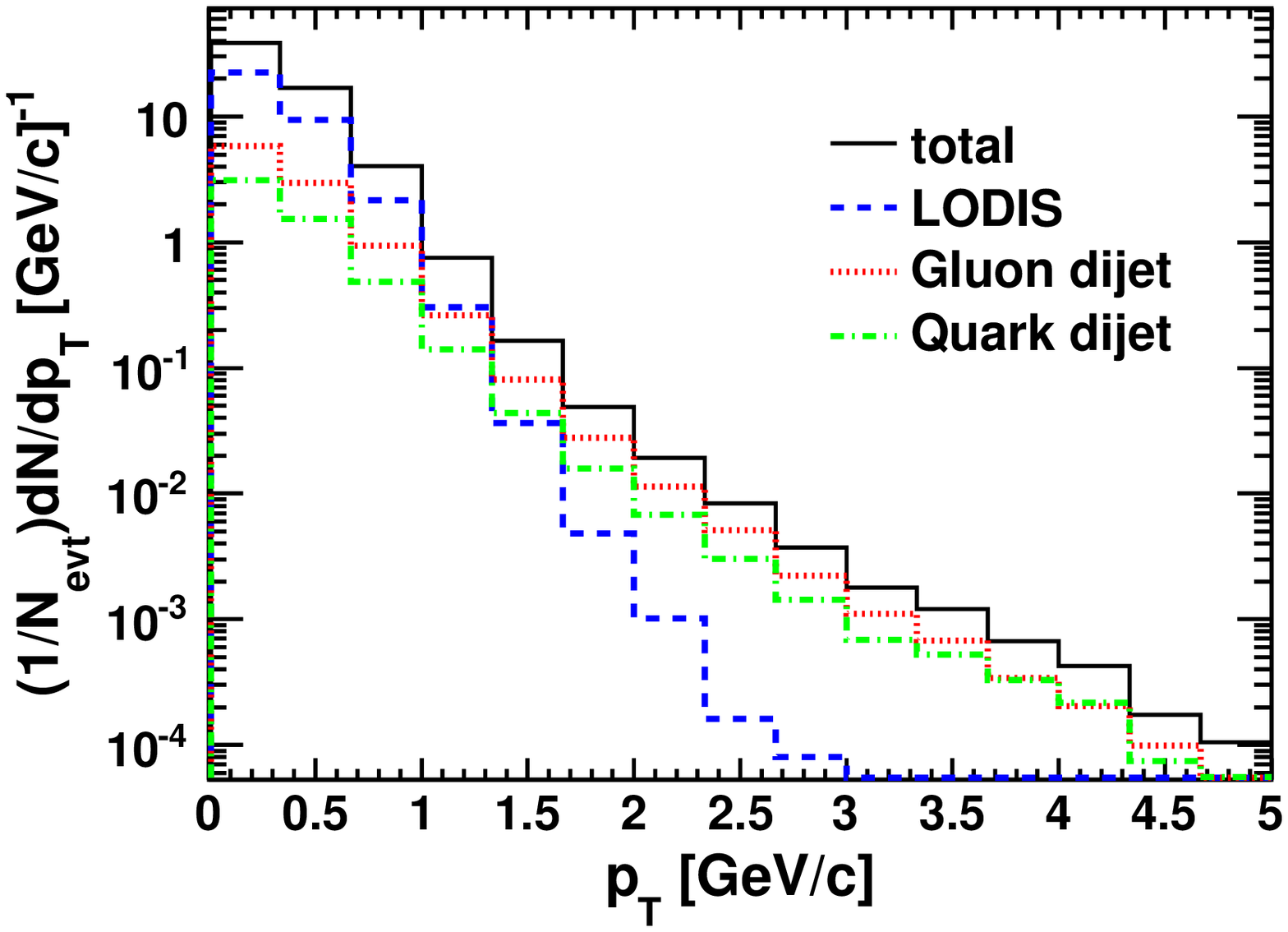}
\includegraphics[width=0.45\textwidth]
{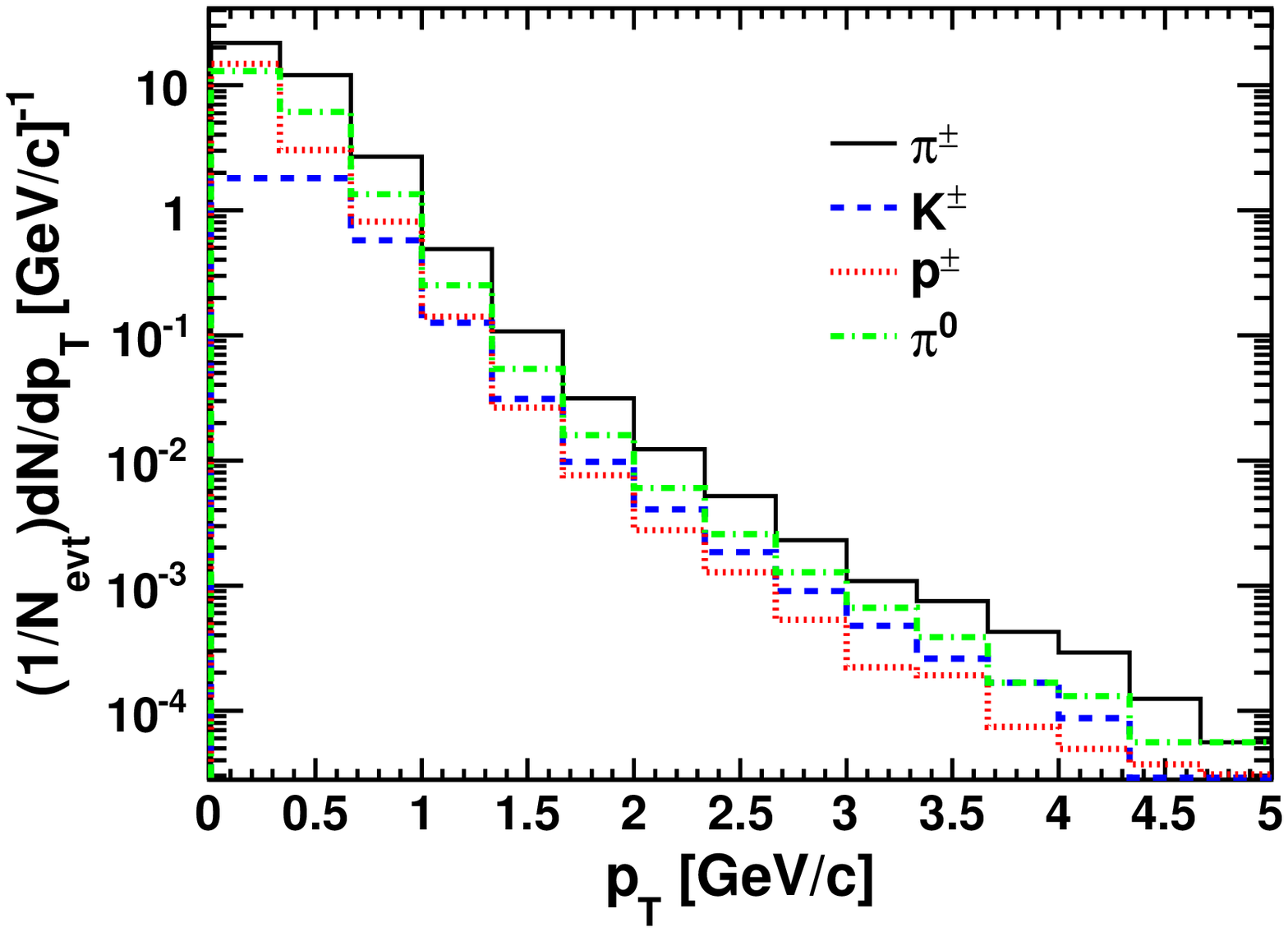}

\end{center} 
\caption[process or PID dependent pt distribution]
{[color online] Particle $p_{T}$ distributions for \ep\ 10 GeV $\times$ 100 GeV collisions with
$1 \, \mathrm{GeV}^{2}  < Q^{2} < 20 \, \mathrm{GeV^{2}}, 0.01 < y < 0.95$. Left: charged particle production from LO DIS, gluon dijets
(PGF and resolved gluon channel) and quark dijets (QCDC and resolved quark channel).
Right: $\pi^{\pm}, K^{\pm}, p^{\pm}$ and $\pi^{0}$ production for all processes.}
\label{fig:PtSpectrum}
\end{figure*}

The present study is based on the planned lepton and nucleon beam energy of 10
GeV $\times$ 100 GeV and 20 GeV $\times$ 100 GeV. The kinematics are constrained
to the main region of interest for dihadron correlation studies,
$1\, \textrm{GeV}^{2}<Q^{2}<20 \, \textrm{GeV}^{2}$ and
$0.01<y<0.95$. Backward electromagnetic calorimetry and the TPC are utilized for
reconstruction of the event kinematics (based on the scattered electron). As shown in
Fig.~\ref{fig:Q2VsxBj}, abundant high-\pt particles will be generated in the
specified kinematic region to make correlated hadron pairs even with a limited
luminosity of 1 $fb^{-1}$. Among those generated high-\pt particles,
Fig.~\ref{fig:PtSpectrum} suggests that gluon dijet processes dominate in the
production of particles with a transverse momentum greater than 2 GeV/$c$ and
charged pions are the major component for the final state particles.

\subsection{Monte Carlo Set Up}
The simulation part of this study is based on the PYTHIA-$6.4$ Monte Carlo program,
with the PDF input from the LHAPDF library~\cite{Whalley:2005nh} and JETSET used for
fragmentation processes. If one defines $\hat{p_{T}}$ as the transverse momentum
of final state partons in the center of mass system of the hard interaction, the
factorization scale $\mu^{2}$ of $2\rightarrow 2$ processes can be expressed
as $\mu^{2} = \hat{p_{T}}^{2} + \frac{1}{2}Q^{2}$.

\begin{figure*} 
\begin{center} 
\subfigure[]{ 
\centering
\includegraphics[width=0.25\textwidth]{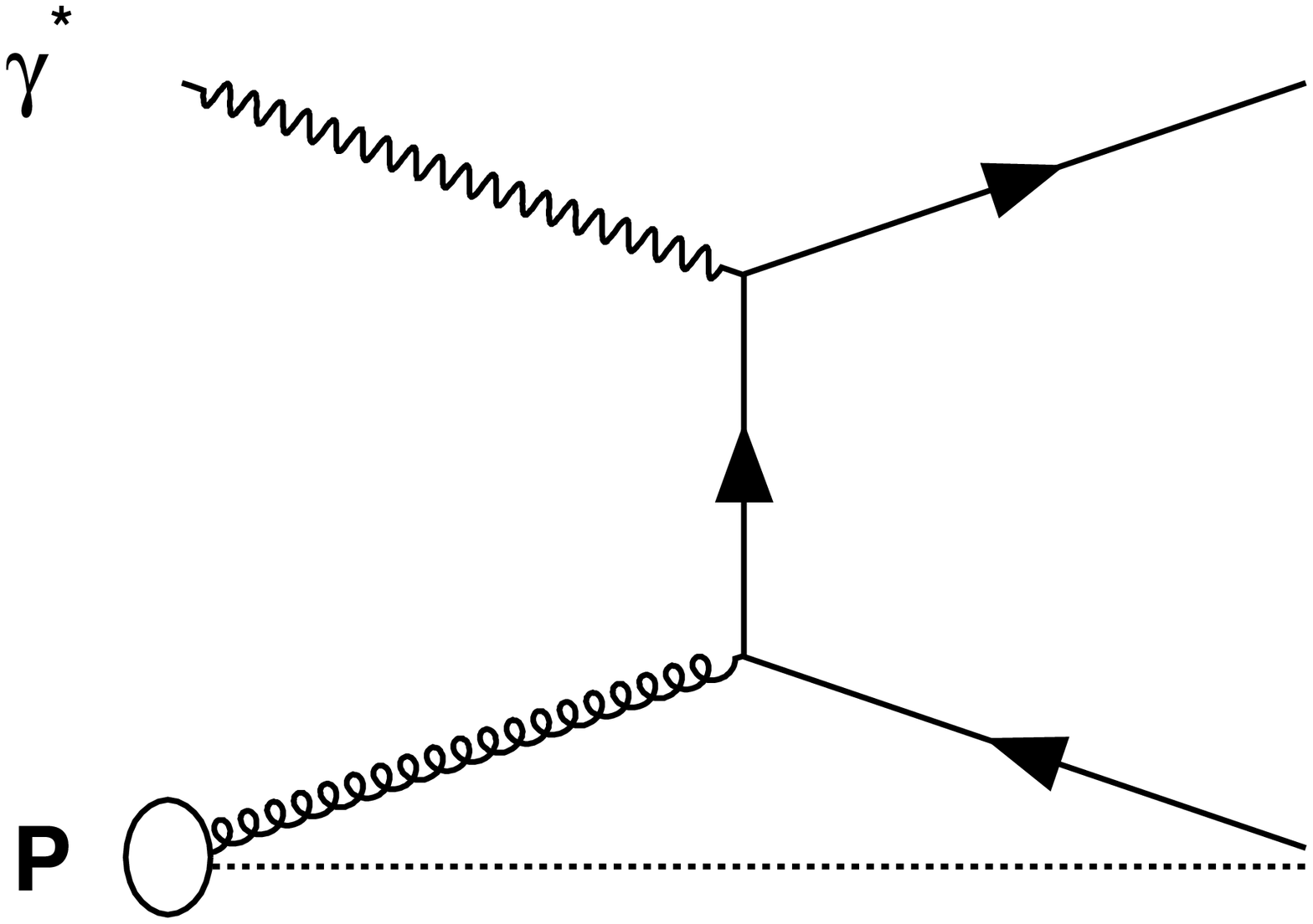} \label{fig:Direct} 
}
\quad 
\subfigure[]{ 
\centering
\includegraphics[width=0.25\textwidth]{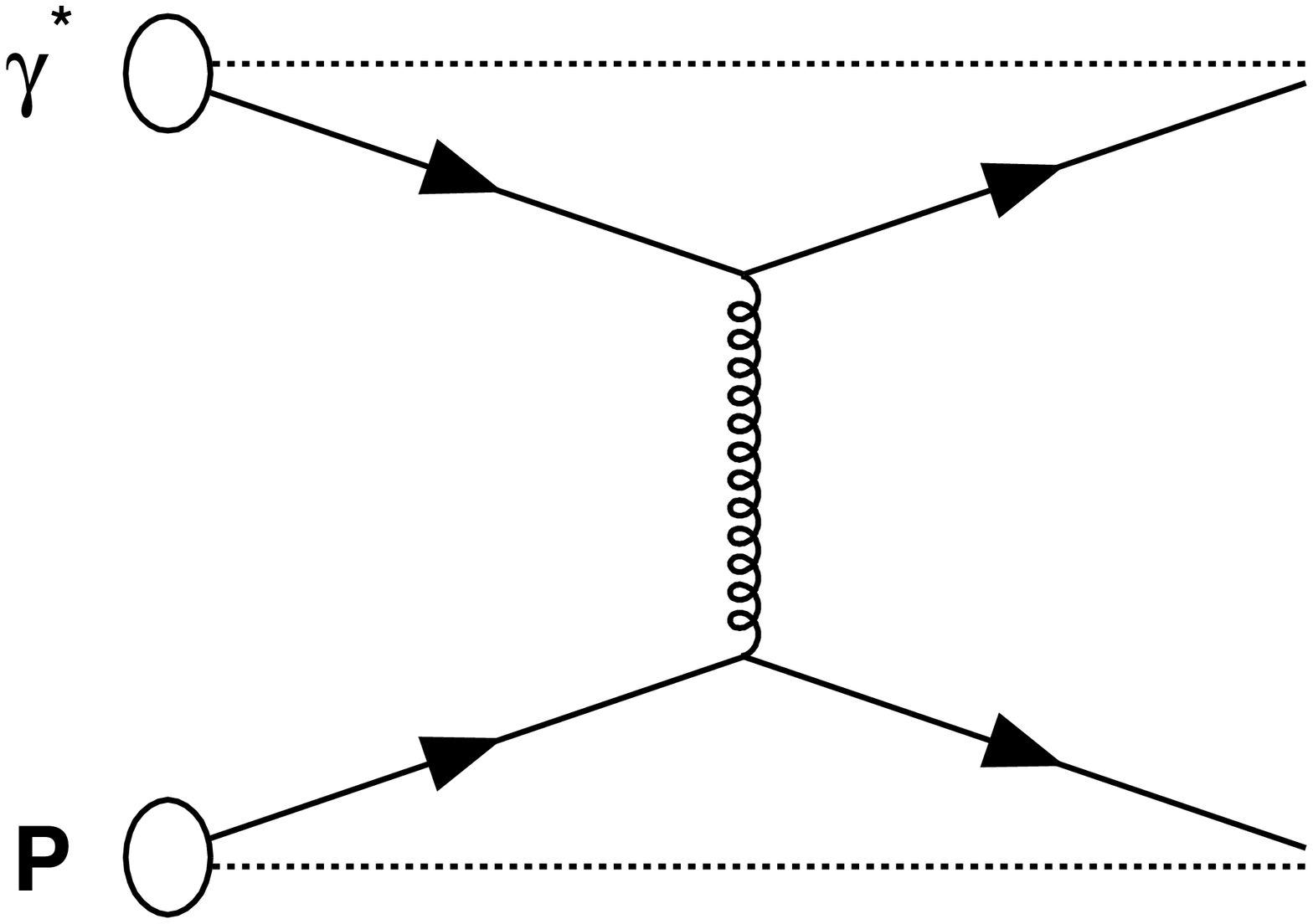} \label{fig:VMD} 
} 
\quad
\subfigure[]{ 
\centering
\includegraphics[width=0.25\textwidth]{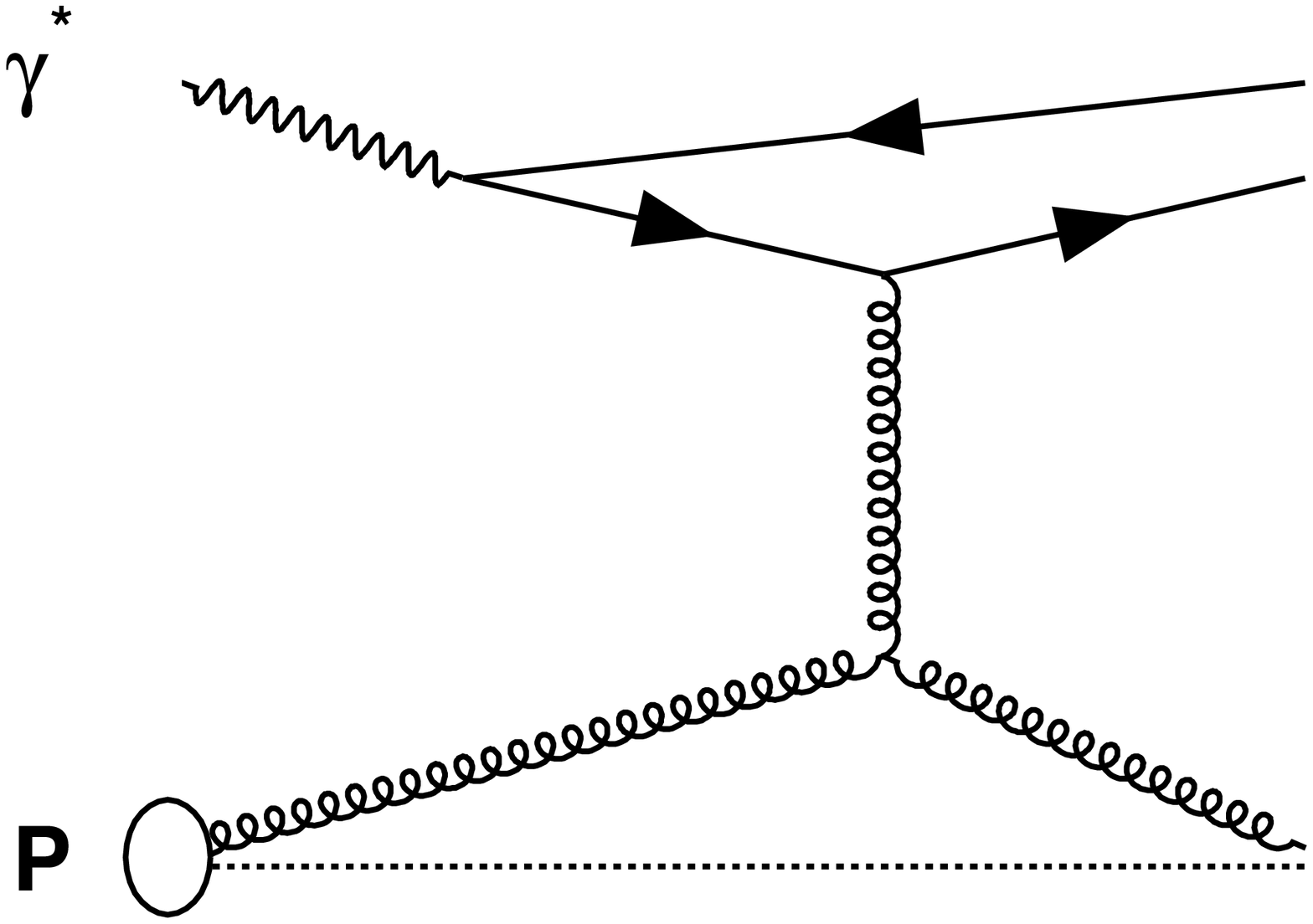} \label{fig:Anomalous} 
}
\caption[PYTIHIA subprocess categories]{Feynman diagrams for different
PYTHIA subprocesses contributing to the hard interaction: (a) direct, (b) VMD,
(c) anomalous. The dotted lines indicate the presence of a spectator. Bubbles
stand for a hadron or hadronic structure.}
\label{fig:PYTHIAFeyn} 
\end{center} 
\end{figure*}

\begin{figure*}
	\begin{center}
	\subfigure[]{
		\centering
		\includegraphics[width=0.25\textwidth]{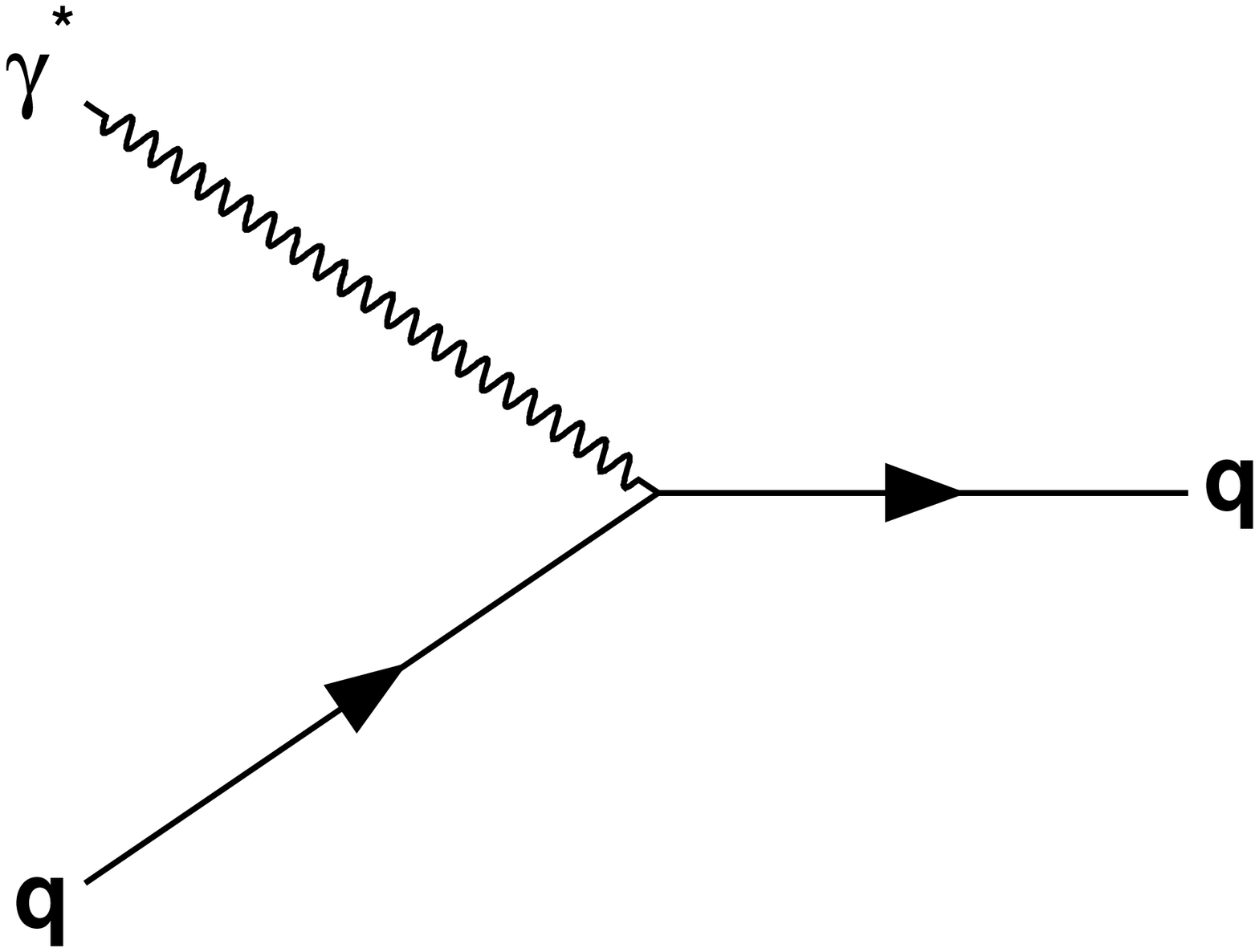}
		\label{fig:LODIS}
	}
	\quad
	\subfigure[]{
		\centering
		\includegraphics[width=0.25\textwidth]{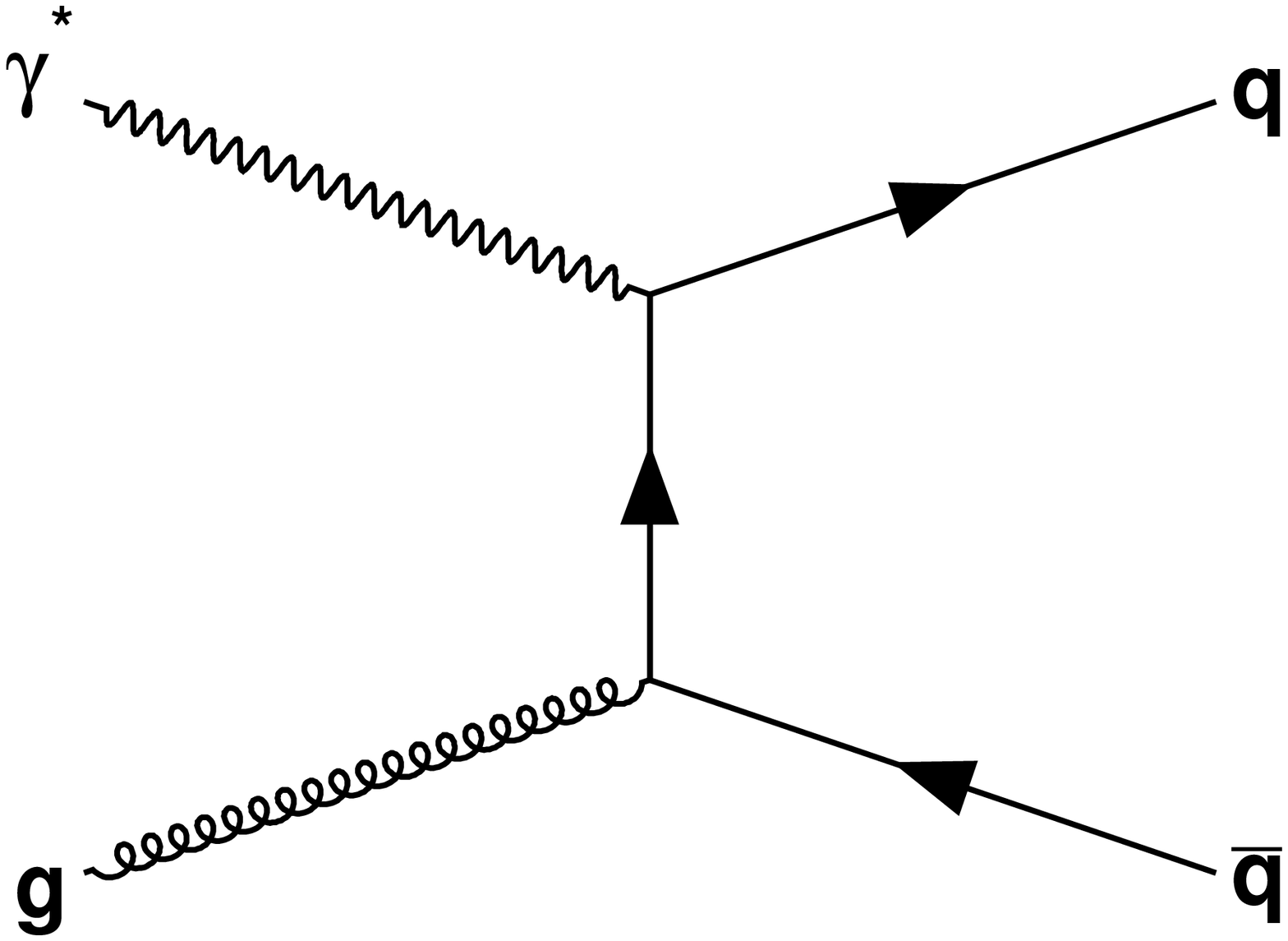}
		\label{fig:PGF}
	}
	\quad
	\subfigure[]{
		\centering
		\includegraphics[width=0.25\textwidth]{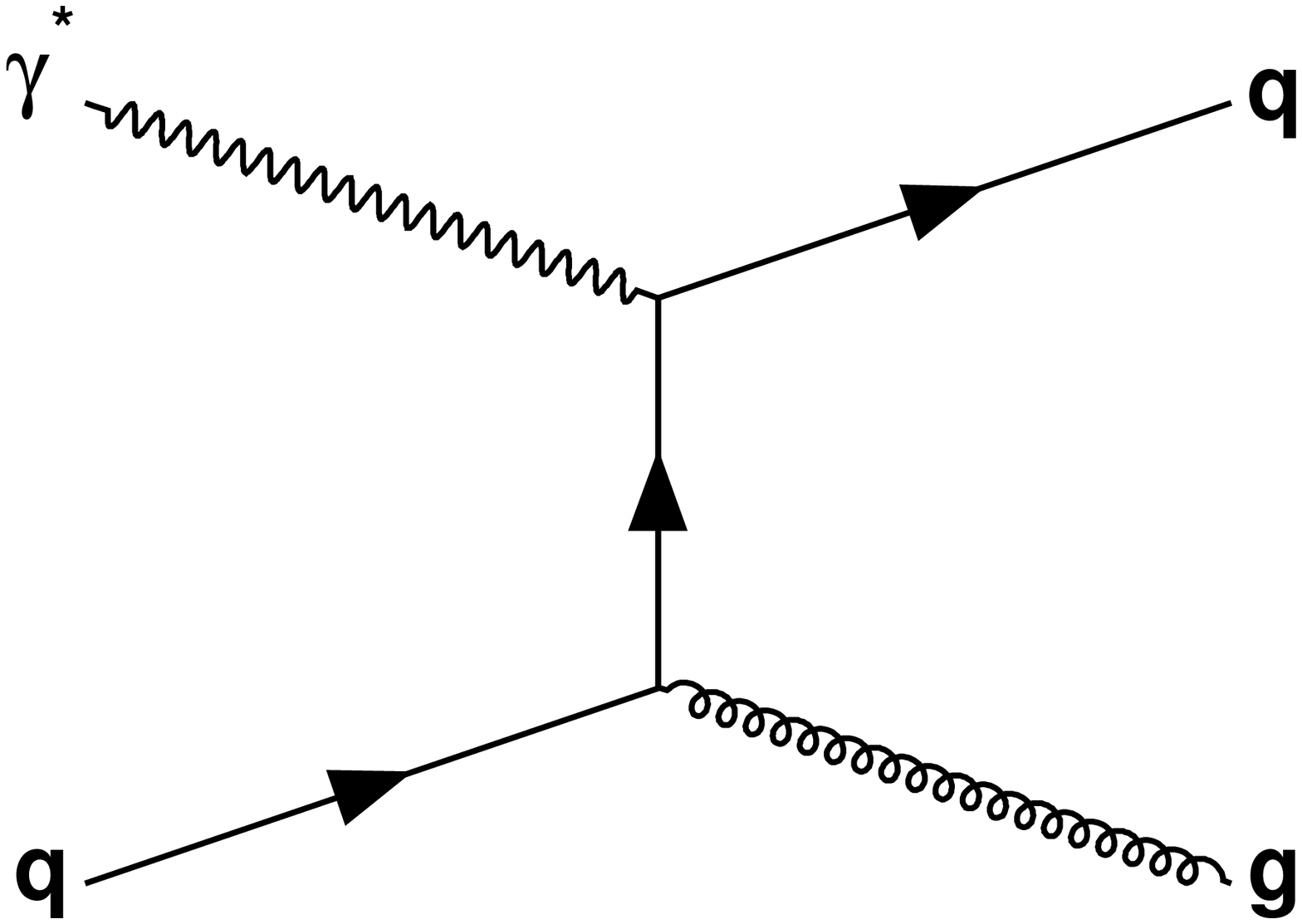}
		\label{fig:QCDC}
	}
	\caption[direct subprocess]{Feynman diagrams for the hard processes 
	based on point-like photons: (a) $\mathcal{O}(\alpha^{0}_{s}) $ LO DIS, (b) 
	Photon-Gluon Fusion (PGF) and (c) QCD Compton scattering (QCDC).}
	\label{fig:DISgraph}
	\end{center}
\end{figure*}

In PYTHIA, depending on the wave function components for the incoming virtual
photon, various subprocesses are divided into three major classes: the direct
processes, the VMD processes and the anomalous processes~\cite{Friberg:2000ra},
as illustrated in Fig.~\ref{fig:PYTHIAFeyn}. The direct photon interacts as a
point-like particle with the partons of the nucleon, while the VMD and anomalous
components interact through their hadronic structure. The VMD component is
characterized by non-perturbative fluctuations of the photon into a $q\bar{q}$
pair existing long enough to evolve into a hadronic state before the subsequent
interaction with the nucleon ~\cite{Bauer:1977iq}. This process can be described
in the VMD model, where the hadronic state is treated as a vector meson (e.g.
$\rho^0$, $\omega$, $\phi$ ) with the same quantum numbers as the photon. These
VMD states can undergo all the soft/hard interactions with the nucleon allowed
in hadronic physics. The large-scale, perturbatively fluctuated photons can be
added as the anomalous photon part in a Generalized VMD (GVMD) model. Same pQCD
$2\rightarrow2$ process can be developed over the VMD or anomalous state of the
virtual photons on the target nucleon, with the difference that parameterized
photon PDFs are used for anomalous photons whereas for the hard VMD components
those vector meson PDFs are involved. Hard VMD and anomalous process are usually
referred to as ``resolved'' process. Resolved photon processes play a significant
part in the production of hard high-$p_{T}$ processes at $Q^{2}\approx0$. The
following hard subprocesses are grouped in the resolved processes category:
$qq\rightarrow qq, \, q\bar q \rightarrow q \bar q, \, q\bar q\rightarrow gg, \,
qg\rightarrow qg, \, gg\rightarrow q\bar q, gg\rightarrow gg$. In the
high-$Q^{2}$ region, direct processes become dominant, Fig.~\ref{fig:DISgraph}
shows the major subprocesses in that category: LO DIS, Photon-Gluon Fusion (PGF)
and QCD Compton (QCDC). As the PGF process is directly sensitive to the gluon
distribution, it is extremely important for DIS dijet productions. The final
conditional yield can be expressed as a superposition from different processes:
\begin{equation} 
C(\Delta\phi)=\Sigma_{i}w_{i}w^{s}_{i}C(\Delta\phi)_{i},
\label{eqn:subprocess} 
\end{equation}
with $w_{i}$ being the statistical weight of every subprocess $i$ involved in
the measurement and $w^{s}_{i}$ is the suppression factor for the subprocess $i$
from saturation. For the quark channels unaffected by saturation $w^{s}_{i}=1$,
while for the gluon channels, suppression of $C(\Delta\phi)$ at away-side is
expected with $w^{s}_{i}<1$.

To simulate \ep\ events the CTEQ6M~\cite{Pumplin:2002vw} PDF in the
$\overline{\textrm{MS}}$ scheme is used. For the \eA\ event sample, the NLO
EPS09 parton distribution functions \cite{Eskola:2009uj} and hard parton energy
loss based on the medium geometry have been applied to account for nuclear
effects in the simulation. Modifications of nuclear PDF have the form
\begin{equation}
f^{A}_{i}(x, Q^{2}) = R^{A}_{i}(x, Q^{2})f^{p}_{i}(x, Q^{2}),
\end{equation} 
where $R^{A}_{i}$ is the nuclear modification factor multiplied on top of the
free proton PDF $f^{p}_{i}(x, Q^{2})$ for a parton of flavor $i$, 
with $f^{p}_{i}(x, Q^{2})$ being the CTEQ6M PDF set. Assuming isospin symmetry
for protons and neutrons, the up/down quark distribution can be
averaged according to the corresponding mass number $A$ and charge number $Z$. 
The hard parton energy loss in the nuclear medium is included
following the parton quenching model (PQM) formalism based on
Ref~\cite{Salgado:2003gb}. In the energy loss picture, the medium effect is
characterized by the so-called transport coefficient, defined as the average
medium-induced squared transverse momentum per unit path length for a hard parton:
\begin{equation} 
\hat{q} =
\left\langle k^{2}_{\perp}\right\rangle _{medium}/\lambda,
\end{equation} 
where $\lambda$ is the mean free path and $k_{\perp}$ represents the
medium-introduced transverse momentum to the hard parton. The characteristic energy loss
scale is set by
\begin{equation} 
\omega_{c} =
\frac{1}{2}\hat{q}L^{2}. 
\end{equation}
$L$ is the medium path length the parton traverses through. It is determined
by the impact parameter at which the hard scattering occurs with respect to 
the geometry of the nucleus described by a Wood-Saxon distribution. Using the above
parameters, one can define the energy loss probability distribution as follows
\begin{equation} 
P(\Delta E; R, \omega_{c}) = p_{0}(R)\delta(\Delta E) + p(\Delta E; R, \omega_{c}),
\end{equation} 
where $p_{0}$ is the probability that the parton experiences no medium induced radiation.

In the phenomenological studies of dihadron correlations, triggering on a hadron
with high-$p_T$ on average selects the most energetic hadron in events with 
back-to-back jets. Correlated hadron pairs reflect two important features of QCD 
dynamics of the hard scattering process.
First, an associated hadron at the near-side allows one to probe the in-medium
QCD evolution of an energetic parton, which can be viewed as the final state
effect with the nuclear medium. Second, an associated hadron at the away-side,
in addition to the primary hard scattering, is sensitive to the initial
transverse momentum that the incoming parton carries.

For pQCD calculations in the collinear factorization framework, the PDFs and fragmentation
functions do not contain any transverse momentum dependence. Therefore, the
transverse momentum of hadrons produced in the final state is given by
$p_{T}=z\hat{p_{T}}$, where $\hat{p_{T}}$ and $p_{T}$ are the transverse
momentum of the parton and hadron respectively. $z$ represents the momentum fraction
of a hadron with respect to its mother parton. This relation should be revised
if the transverse momentum is allowed in both the PDFs and fragmentation
functions.

Transverse motion of partons inside hadrons can be effectively included by
assuming that the intrinsic \kt follows a Gaussian distribution. Similarly, the
transverse momentum enhancement  $p_{T}^{\textrm{frag}}$ with respect to the jet
direction during hadronization can also be approximated by a Gaussian
distribution. The intrinsic \kt and fragmentation $p_{T}^{\textrm{frag}}$ now
both contribute to the transverse momentum of final state hadrons, which can be
written as $p_{T} =z(k_{T}+\hat{p_{T}})+p_{T}^{\textrm{frag}}$. We follow the
common practice to set the Gaussian width to $0.4$ GeV for both intrinsic $k_{T}$ 
and $p_{T}^{\textrm{frag}}$ distributions in the simulations.

Besides all the above effects, additional soft gluon radiations, normally 
characterized as a parton shower can also modify the final transverse
momentum, thereby impacting the dihadron correlations. In perturbative QCD
calculations the parton shower are computed in terms of Sudakov
form factors.

\begin{figure}
\begin{center}
\includegraphics[width=0.5\textwidth]
{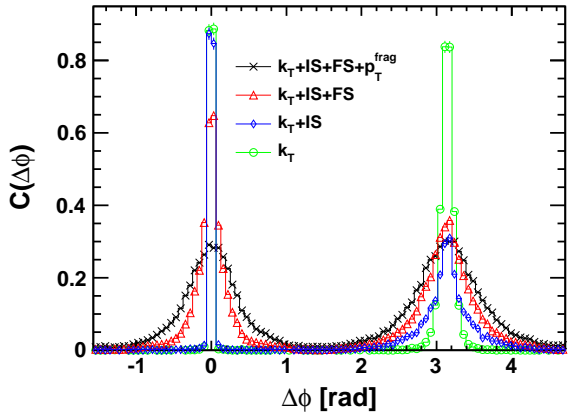}
\end{center} 
\caption[effects in Monte Carlo on azimuthal diahdron correlation]
{[color online] Comparison of dihadron correlation due to different physical inputs, such as 
intrinsic $k_T$, initial state parton shower (IS), final state parton shower plus 
resonance decay (FS) and $p_T$ broadening in fragmentation processes.
The \ep\ data are for charged hadrons with a beam energy of 20 GeV $\times$ 100 GeV 
with $1.0 \, \mathrm{GeV}^{2}  < Q^{2} < 1.5 \, \textrm{GeV}^2, 0.65 < y < 0.75, p_{T}^{trig} > 2 \,
\mathrm{GeV/}c, 1 \, \mathrm{GeV/}c < p_{T}^{assoc} < p_{T}^{trig}, 0.2 <
z_{h}^{trig}, z_{h}^{assoc} < 0.4$. }
\label{fig:dihadron_effects}
\end{figure}

Fig.~\ref{fig:dihadron_effects} shows an illustration of all the possible
effects available in the Monte Carlo in the simulation of the azimuthal correlation
function. The open circles illustrate the dihadron correlation with only
intrinsic $k_T$ in the initial parton distribution. It is understandable that the
correlation function is strongly peaked at $\Delta \phi =0, \pi$ for this setting.
Now the other effects are turned on one-by-one according to the order of their
occurrences in physical processes. When the initial state (IS) parton shower
is added into the simulation, as shown by the open diamonds, the away-side
correlation is significantly reduced since it is very sensitive to the momentum
imbalance of the dijet system, while the near-side correlation is almost
unmodified. Next, we turn on the final state (FS) parton shower for the scattered 
parton before the fragmentation process occurs. We find that both the
near-side and away-side peaks are broadened, as illustrated by the empty triangles,
due to soft radiation and particle decay in the fragmentation. 
Lastly, we add transverse momentum dependence into the fragmentation 
function, labeled as $p_T^{\textrm{frag}}$, and obtain the crossings, which indicate further broadening of both peaks.

\begin{figure}
\begin{center}
\includegraphics[width=0.5\textwidth]
{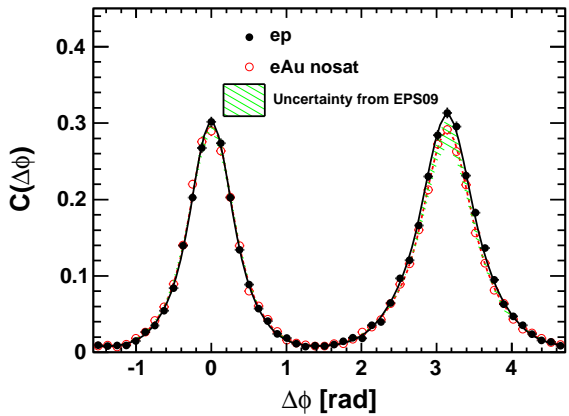} 
\end{center} 
\caption[Monte Carlo result of diahdron correlation]{[color online] Simulated data points for particle correlations
for charged hadrons in \ep\ and \eAu\ collisions with beam energies of 20 GeV $\times$ 100
GeV and $1.0 \, \mathrm{GeV}^{2}  < Q^{2} < 1.5 \, \mathrm{GeV}^{2} , 0.65 < y < 0.75, p_{T}^{trig} > 2 \, \mathrm{GeV/}c, 1
\, \mathrm{GeV/}c < p_{T}^{assoc} < p_{T}^{trig}, 0.25 < z_{h}^{trig}, z_{h}^{assoc} <
0.35$. Lines are the fit for \ep\ or \eA\ points. The shaded band shows the uncertainty due to the EPS09 nuclear PDFs.
\label{fig:dihadron_base}}
\end{figure}

For our model of \eA\ implemented in PYTHIA, the effects due to energy loss
in the cold nuclear medium are expected to be weak, because fast moving partons are likely to
fragment outside the nucleus in the considered kinematic regions. Considering
that the nuclear PDF also has little impact on the $p_{T}$ imbalance of dijets,
it comes as no surprise to see very little change from \ep\ to \eA\ in the
simulation, as shown in Fig.~\ref{fig:dihadron_base}.
\begingroup
\squeezetable
\begin{table} 
\centering 
\caption[table for effects on azimuthal correlation function]{Relative 
Root Mean Square (RMS) for the $\Delta\phi$ distribution from \ep\ collisions including
different effects influencing the width of the near and away side peak compared
to the baseline RMS with all the effects included (bottom row).}
\label{tab:azimuRMS} 
\begin{tabular}{ l  c  c  } \hline \hline
		& Near-side $\Delta\phi$ RMS & Away-side $\Delta\phi$ RMS \\ \hline
\kt		  	&  0.21  &  0.25   \\  
\kt+ IS     &  0.30  &  0.72   \\  
\kt+ IS + FS    & 0.65  &  0.81   \\  
\kt+ IS + FS + $p_T^{\textrm{frag}}$    &  1.00  & 1.00   \\ \hline \hline
\end{tabular} 
\end{table}
\endgroup
Table~\ref{tab:azimuRMS} is a reference for different effects on the relative root mean
square (RMS) deviation of the near/away side azimuthal correlation function,
from which we can clearly draw the conclusion that initial-state parton showers
dominate the away-side peak of the correlation function, while the near-side peak 
is mainly controlled by final-state effects such as final-state parton showers, 
fragmentation \pt and possible resonance decays in the fragmentaion.

As the saturation physics discussed above is mainly about the gluon dynamics, in
order to be able to consistently compare with the theoretical dihadron cross
section in Sec.~\ref{sec:theory}, we need to include gluon dijet channels from
PGF and gluon-initiated resolved process in the comparison. However, as the
measured observable in the real experiment is a mixture of different process, as
illustrated in Eq.~(\ref{eqn:subprocess}), we have to know how significant the
signal from gluon saturation manifests itself in a mixed event sample. From the
saturation-based predictions, a sizeable suppression of the away-side peak from \ep\
to \eA\ is expected.

In the meanwhile, it is crucial to point out that parton showers suppress the
away-side peak of the dihadron correlation function just like saturation does.
However, currently it is still unclear how the parton shower effect is modified
in the nuclear medium, without which it is hard to draw any definite conclusions
about the saturation effects, as parton showers and saturation effects are always
entangled. Nevertheless, thanks to the large kinematic coverage of eRHIC, one
can explore the nuclear dependence of parton showers outside the saturation
region by measuring dihadron correlations for different nuclei in the high
$Q^2$ regime. This kinematic regime has a significant phase space for parton
showers for this observable. More importantly, the measurement of dihadron
correlations gives the opportunity to use the near-side peak of the correlation
function as a reference to study the nuclear medium effects on parton showers as
the saturation effects only manifest themselves in the away-side peak, as shown
in Fig.~\ref{fig:dihadron_effects}.

\begin{figure} 
\begin{center}
\includegraphics[width=0.5\textwidth]{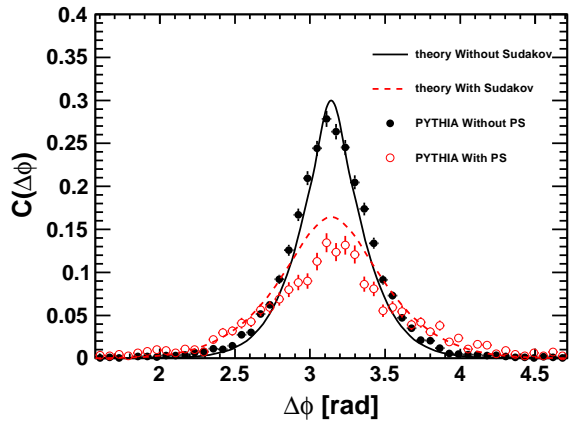} 
\end{center} 
\caption[comparison of ep from PYTHIA and theory with Sudakov]
{[color online] \pion $\Delta\phi$ correlation comparing PYTHIA and theoretical saturation calculations
for \ep\ 10 GeV $\times$ 100 GeV for events from PGF and resolved gluon channel subprocesses
 at $1.0 \, \mathrm{GeV}^{2} < Q^{2} < 2.0 \, \mathrm{GeV}^{2}$, \, $0.65 < y < 0.75, \, p_{T}^{trig} > 2 \,
\mathrm{GeV/}c, 1 \, \mathrm{GeV/}c < p_{T}^{assoc} < p_{T}^{trig}, \,0.25 <
z_{h}^{trig}, z_{h}^{assoc} < 0.35$. The solid and dashed curves show theoretical predictions including 
saturation effects for \ep\ without and with Sudakov factor, respectively. 
The filled and empty circles illustrate PYTHIA simulations for \ep\ without and with parton showers. }
\label{fig:epCompareWithSud} 
\end{figure}

In the saturation formalism, the parton shower contribution is effectively cast
into the Sudakov factor for the DIS dijet process at small $x$. To illustrate this
point, Fig.~\ref{fig:epCompareWithSud} shows the correlation function simulated
with and without parton showers, compared to the corresponding theoretical
predictions with and without Sudakov effects. The filled circles represent the PYTHIA
simulation for \ep\ without parton showers, and they agree very well with the solid
line from the theoretical prediction including saturation effects, but excluding
Sudakov effects. The comparison (empty circles and dashed line) between simulated PYTHIA \ep\
data including parton showers and the theoretical predictions with
saturation plus Sudakov effects is also good, especially considering the model
uncertainties. Thus, the agreement in \ep\ collisions enables one to estimate the
nuclear medium effects on parton showers in the theoretical predictions for
saturation including Sudakov effects.

Since the saturation effect decouples from hadronization, it does not
depend on which specific particle type being detected. Although the theoretical
prediction is made for $\pi^{0}$, the suppression factor from \ep\ to \eA\
still holds for other different final state particles. In the next section, the
significance for the suppression of gluon saturation will be shown for the
charged hadrons $C(\Delta\phi)$ observable with limited statistics and
expected background estimation.

\subsection{Monte Carlo Results and Uncertainties} 

\begin{figure*}
\begin{center}
\includegraphics[width=0.8\textwidth]{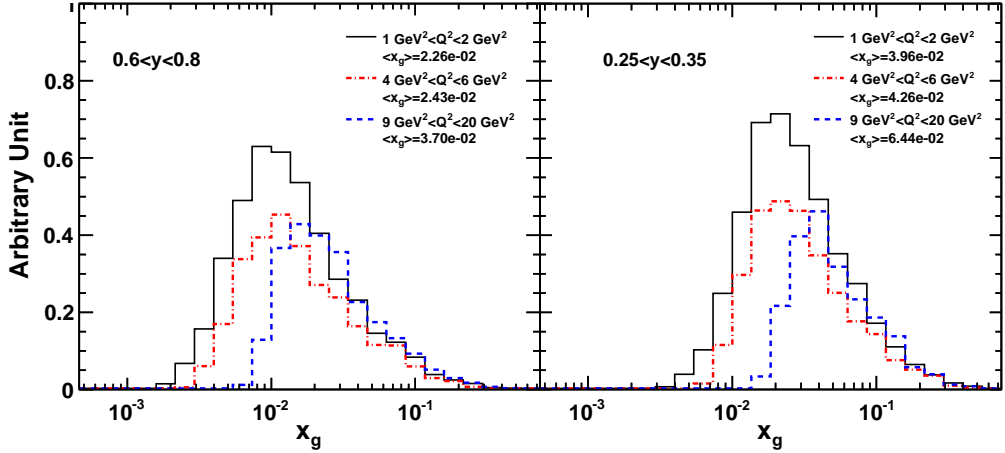} 
\end{center} 
\caption[$x_{g}$ coverage]{[color online] $x_{g}$ distributions in various kinematics bins
probed by the correlated hadron pairs in the PGF process for \ep\ 10 GeV
$\times$ 100 GeV. }
\label{fig:xgCover} 
\end{figure*}

In order to guarantee the validity of perturbative calculations and avoid the
kinematic regime of quasi-real photo-production, a cut of $Q^{2}>1
\, \mathrm{GeV}^{2}$ is generally made. On the other hand, probing the
saturation dynamics requires to probe the dense region, which means that one
needs to go to low $x$ and low-to-moderate $Q^{2}$ in the pursuit of
saturation effects at a certain center of mass energy. A cut in transverse
momentum of the charged hadron pairs is usually performed to pick particles from
hard interactions. A cut on $z_{h}$ is also imposed to reject particles from the
target remnants. Thus, a typical cut to select dihadron pairs from hard parton
scatterings is: $p_{T}^{trig}>2$ GeV/$c$, $1 \, \mathrm{GeV/}c
<p_{T}^{assoc}<p_{T}^{trig}, 0.2<z_{h}^{trig},z_{h}^{assoc}<0.4$. To further explore the
transition behavior in and out of the saturation region, we use three $Q^{2}$ bins:
$1\, \textrm{GeV}^{2}<Q^{2}<2 \, \mathrm{GeV}^{2}$, $3\, \textrm{GeV}^{2}<Q^{2}<5 \,
\mathrm{GeV}^{2}$ and $9\, \textrm{GeV}^{2}<Q^{2}<20 \, \mathrm{GeV}^{2}$; and two $y$ bins:
$0.25<y<0.35$ and $0.6<y<0.8$. To study saturation physics using varying heavy
ion beams, we focus on the $1\, \textrm{GeV}^{2}<Q^{2}<2 \, \mathrm{GeV}^{2}$, $0.6<y<0.8$
bin, while the $9\, \textrm{GeV}^{2}<Q^{2}<20 \, \mathrm{GeV}^{2}$ or $0.25<y<0.35$ bins serve
as a reference for the behaviour without saturation effects. To pin down the
nuclear dependence of parton showers, we compare the correlation between \ep\
and \eA\ collisions for bins without saturation effects.
\begin{figure}
\begin{center}
\includegraphics[width=0.5\textwidth]{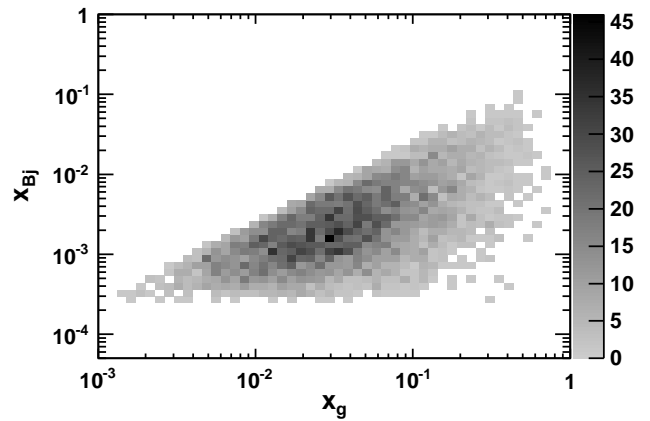}
\end{center}
\caption{The correlation of $x_{Bj}$ vs $x_{g}$ for the PGF process for \ep\ at 
10 GeV $\times$ 100 GeV with $0.01<y<0.95$, $1\, \textrm{GeV}^{2}<Q^{2}<20 \, \mathrm{GeV^{2}}$. A relatively 
broad correlation between these two variables is observed. }
\label{fig:xbjVsxg}
\end{figure}
Because the saturation scale $Q_{s}$ varies with the gluon momentum fraction
$x_{g}$, it is important to have access to $x_{g}$.
Fig.~\ref{fig:xbjVsxg} shows how, by utilizing $x_{Bj}$, one can
effectively constrain the underlying $x_{g}$ distribution. Although this is only
a broad correlation, it is demonstrated in Fig.~\ref{fig:xgCover} that
the typical $x_{g}$ at a given $x_{Bj}$ is constrained to a certain magnitude and can be used to
separate the saturation region from the non-saturated region.

\begin{figure*}
\begin{center}
\includegraphics[width=0.8\textwidth]{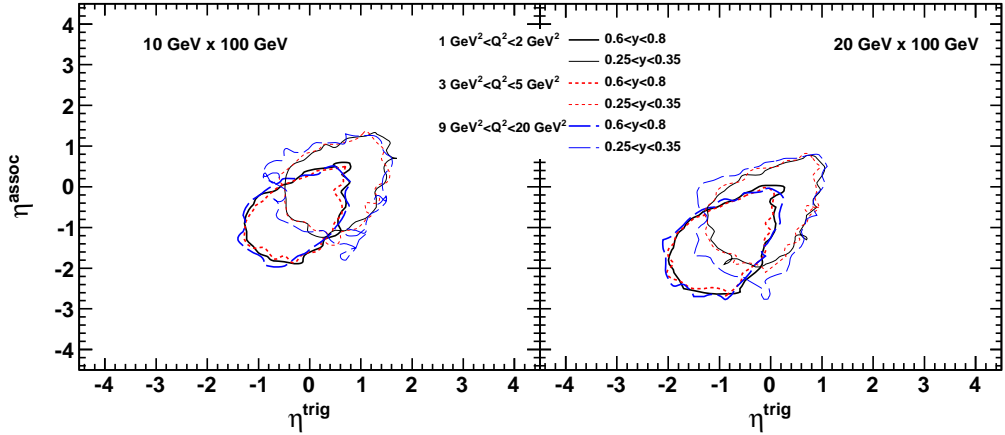}
\end{center} 
\caption[dihadron pair $\eta$ distribution]{[color online] The contours show the
$\eta$ regions covered by the correlated dihadron pairs: for 10 GeV $\times$ 100
GeV and 20 GeV $\times$ 100 GeV. Thick lines mark out the region for $0.6<y<0.8$
while thin lines for $0.25<y<0.35$.}
\label{fig:pairEta} 
\end{figure*}

Fig.~\ref{fig:pairEta} shows the $\eta$ distribution of the trigger
particle and the correlated particle in the aforementioned kinematic
bins at 10 GeV $\times$ 100 GeV and 20 GeV $\times$ 100 GeV. Clearly, with a
charged particle acceptance spanning $-4.5<\eta<4.5$, both trigger and
associate particles in our kinematics binning scheme can be fully accepted by
the detector.

\begin{figure*}[hbt]
\begin{center}
\includegraphics[width=0.8\textwidth]{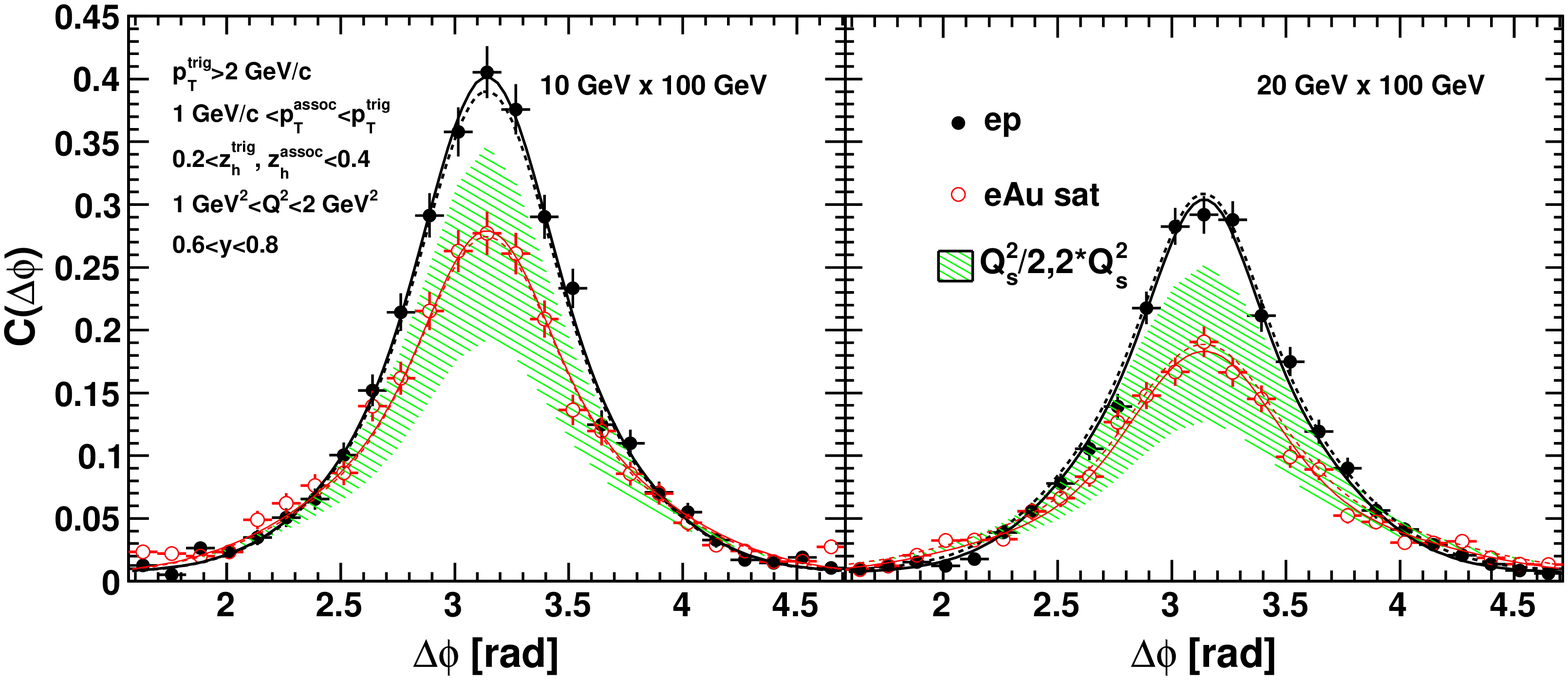} 
\end{center} 
\caption[correlation function with saturation scale uncertainty]{[color online] The correlation
function at $1\, \textrm{GeV}^{2}<Q^{2}<2 \, \textrm{GeV}^{2}$, $0.6<y<0.8$ for an  
integrated luminosity of $1 \, fb^{-1}$. The \ep\ result comes from PYTHIA simulations.
The $e+$Au results are a combination of simulations from a saturation-based model plus modified
PYTHIA simulations. The suppression factor uncertainty was estimated by varying
$Q_{s}^{2}$ by a factor of 0.5 and 2. Sudakov resummation has also been incorporated
for \eAu. The solid lines represent a fit for the simulated pseudo-data
including detector effects; the dashed line excludes detector effects.}
\label{fig:correUncertainSud}
\end{figure*}

In Fig.~\ref{fig:correUncertainSud} we compare the strength of the
coincidence probability based on a theoretical saturation model prediction for
the away-side for \ep\ and $e+$Au. The filled circles in
Fig.~\ref{fig:correUncertainSud} are simulated with PYTHIA
for \ep\ collisions, including detector smearing and acceptance effects.
The projected $e+$Au saturated correlation function (open circles) is
obtained by multiplying the \ep\ histogram with the suppression factor
$w^{s}_{i}=C(\Delta\phi)_{eAu}/C(\Delta\phi)_{ep}$ including Sudakov effects extracted
from Fig.~\ref{fig:dihadron_theory_sud}. This suppression factor can only be applied to
dijet channels involving gluons; namely PGF and resolved $qg\rightarrow qg$,
$gg\rightarrow gg$, $gg\rightarrow q\bar q$ subprocesses. The other quark
initiated subprocesses have been simulated with PYTHIA using the non-saturated
\eA\ model including nuclear PDFs and final-state energy loss. The uncertainties
represent the statistical precision from an integrated luminosity of $1 \,
fb^{-1}$. The solid (dashed) lines in Fig.~\ref{fig:correUncertainSud} represent
fits to the simulated data points with (without) detector effects included in
the simulation.

Since to date there is no exact knowledge of the saturation scale, the
uncertainty in the suppression factor is estimated by varying the saturation
scale by a factor of 0.5 and 2. The resulting uncertainty bands are depicted in
Fig.~\ref{fig:correUncertainSud}. The suppression of the away-side peak remains
significant even with this additional uncertainty compared to the $e+$Au curve
shown in Fig.~\ref{fig:dihadron_base} accounting for nuclear effects in the
parton distribution functions, energy loss effects and the resonance decay. In
summary, the suppression effects on dihadron correlations due to saturation can
be clearly discriminated from effects based on classical nuclear medium
modifications with a well-designed EIC machine.

\section{Summary}\label{sec:summary}

Through detailed analysis, we have shown the capability of a proposed EIC
to perform dihadron correlation measurements. It
is proven that the acceptance of the dedicated detector is wide enough to
collect all the trigger particles as well as the associated particles used in our
studies. Moreover, the onset of the projected saturation region is well covered by
the eRHIC energy regime. It will clearly be possible to do a high
precision measurement of the correlation function for dihadron production with
different nuclear beams at the proposed EIC.

In this study we also describe how this dihadron correlation function is
calculated in a saturation/CGC formalism, and provide predictions for this
measurement with or without saturation effects taken into consideration. It is
straightforward to see that a strong suppression of the away-side peak of the
correlation function is expected from saturation effects, and detector effects are
negligible on this observable. Suppression effects due to leading-twist
shadowing are significantly smaller. Therefore, the observation of such
a suppression in the dihadron correlation function measured at an EIC will be a
strong experimental evidence for the existence of gluon saturation.

Dihadron measurements at an EIC are also vital and intriguing in
that they will directly measure for the first time the behavior of the Weizs\"{a}cker-Williams
gluon distribution, about which we still know very little, and which we
can hardly extract from other measurements. The knowledge of how parton showers
behave in a nuclear medium is indispensable in obtaining a valid conclusion for the
above discussions. With the Sudakov resummation performed in the saturation
formalism, this nuclear modification of parton showers in DIS dijet process is
found to be very small at leading order. Nevertheless, there might be some
nuclear dependence in the Sudakov factor at higher orders or in the
non-perturbative part~\cite{Kang:2012am}. An EIC will also permit unique measurements
that will give a definite answer to this question. With the wide kinematic
reach and different nuclear beams, an EIC is capable of measuring the A
dependence of parton showers. This would require quantitative measurements of
the modification of the near-side peak of the correlation functions in a 
nucleus environment and in kinematic regions where parton showers dominate.

In conclusion, the proposed high-luminosity, high-energy Electron-Ion Collider,
together with the designed detector, can provide an ideal apparatus to study
gluon saturation with high precision through the measurement of 
the dihadron correlation function.

\begin{acknowledgments}
This work was supported in part by the NSFC (11375071), the National Basic Research 
Program of China (2013CB837803), and the Basic Research Program of CCNU (CCNU13F026). 
E.C.A. and J.H.L. acknowledge support by the U.S. Department of Energy
under contract number DE-AC02-98CH10886. 
\end{acknowledgments}


\end{document}